\documentclass[twocolumn]{aastex63}
\usepackage[]{natbib}
\usepackage{graphicx, amsmath}
\usepackage[caption=false]{subfig}
\usepackage{paralist}
\usepackage{hyperref}
\hypersetup{
    colorlinks=true,
    linkcolor=blue,
    filecolor=magenta,      
    urlcolor=cyan,
}
\def\etal{{et~al.\null}}
\def\eg{e.g.,}
\def\ie{i.e.,\ }

\newcommand{\Hb}{H$\beta$}
\newcommand{\OII}{[\ion{O}{2}]}
\newcommand{\OIII}{[\ion{O}{3}]}
\newcommand{\NeIII}{[\ion{Ne}{3}]}

\newcommand{\mcsed}{\texttt{MCSED}}
\newcommand{\cloudy}{C{\footnotesize LOUDY}}
\newcommand{\threedhst}{\mbox{3D-HST}}

\begin{document}

\title{\mcsed : A flexible spectral energy distribution fitting code and its application to $\lowercase{z}\sim2$ emission-line galaxies}
\shorttitle{MCSED}

\author{William P. Bowman}
\affiliation{Department of Astronomy \& Astrophysics, The Pennsylvania
State University, University Park, PA 16802}
\affiliation{Institute for Gravitation and the Cosmos, The Pennsylvania State University, University Park, PA 16802}

\author{Gregory R. Zeimann}
\affiliation{Hobby Eberly Telescope, University of Texas, Austin, Austin, TX, 78712, USA}

\author{Gautam Nagaraj}
\affiliation{Department of Astronomy \& Astrophysics, The Pennsylvania
State University, University Park, PA 16802}
\affiliation{Institute for Gravitation and the Cosmos, The Pennsylvania State University, University Park, PA 16802}

\author{Robin Ciardullo}
\affiliation{Department of Astronomy \& Astrophysics, The Pennsylvania
State University, University Park, PA 16802}
\affiliation{Institute for Gravitation and the Cosmos, The Pennsylvania State University, University Park, PA 16802}

\author{Caryl Gronwall}
\affiliation{Department of Astronomy \& Astrophysics, The Pennsylvania
State University, University Park, PA 16802}
\affiliation{Institute for Gravitation and the Cosmos, The Pennsylvania State University, University Park, PA 16802}

\author{Adam P. McCarron}
\affiliation{Department of Astronomy \& Astrophysics, The Pennsylvania
State University, University Park, PA 16802}

\author{Laurel H. Weiss}
\affiliation{Department of Astronomy \& Astrophysics, The Pennsylvania
State University, University Park, PA 16802}

\author{Mallory Molina}
\affiliation{Department of Astronomy \& Astrophysics, The Pennsylvania
State University, University Park, PA 16802}
\affiliation{Institute for Gravitation and the Cosmos, The Pennsylvania State University, University Park, PA 16802}
\affiliation{Department of Physics, Montana State University, Bozeman, MT 59717}

\author{Alexander Belles}
\affiliation{Department of Astronomy \& Astrophysics, The Pennsylvania
State University, University Park, PA 16802}
\affiliation{Institute for Gravitation and the Cosmos, The Pennsylvania State University, University Park, PA 16802}

\author{Donald P. Schneider}
\affiliation{Department of Astronomy \& Astrophysics, The Pennsylvania
State University, University Park, PA 16802}
\affiliation{Institute for Gravitation and the Cosmos, The Pennsylvania State University, University Park, PA 16802}

\begin{abstract}
We present \mcsed, a new spectral energy distribution (SED)-fitting code, which mates flexible stellar evolution calculations with the Markov Chain Monte Carlo algorithms of the software package \texttt{emcee}. \mcsed\ takes  broad, intermediate, and narrow-band photometry, emission-line fluxes, and/or absorption line spectral indices, and returns probability distributions and co-variance plots for all model parameters. \mcsed\ includes a variety of dust attenuation curves with parameters for varying the UV slopes and bump strengths, a prescription for continuum and PAH emission from dust, models for continuum and line emission from ionized gas, options for fixed and variable stellar metallicity, and a selection of star formation rate (SFR) histories.  The code is well-suited for exploring parameter inter-dependencies in sets of galaxies with known redshifts, for which there is multi-band photometry and/or spectroscopy.  We apply \mcsed\ to a sample of $\sim2000$ $1.90<z<2.35$ galaxies in the five CANDELS fields, which were selected via their strong \OIII\ $\lambda5007$ emission, and explore the systematic behavior of their SEDs.  We find the galaxies become redder with stellar mass, due to both increasing internal attenuation and a greater population of older stars.  The slope of the UV extinction curve also changes with stellar mass, and at least some galaxies exhibit an extinction excess at 2175~\AA\null.  Finally, we demonstrate that below $M\lesssim10^9\,M_{\odot}$), the shape of the star-forming galaxy main sequence is highly dependent on the galaxies' assumed SFR history, as calculations which assume a constant SFR produce stellar masses that are $\sim1$~dex smaller than those found using more realistic SFR histories.

\end{abstract}

\keywords{galaxies: formation --- galaxies: evolution -- galaxies:  stellar content --- dust, extinction --- cosmology: observations}

\section{Introduction}
\label{sec:Intro}
Our primary source of information for understanding the evolutionary state of a distant stellar population is its integrated light as a function of wavelength, i.e., its spectral energy distribution (SED\null).  Indeed, a galaxy's ultraviolet (UV) through far-infrared (far-IR) SED contains insights into its current star-formation rate, past history of star-formation, and present day stellar mass, dust content, and chemistry. Consequently, much of what is known about the galaxies of the $z \gtrsim 2$ universe is based on SED fits, and numerous programs now exist to perform this analysis \citep[e.g.,][]{GalMC, BayeSED, BEAGLE, prospector, FIREFLY}.

SED modeling is effective because different physical processes leave their imprint on different regions of the electromagnetic spectrum.  For example, estimates of the present-day star-formation rate and dust attenuation are obtained primarily from the UV, stellar masses and metallicities are derived largely from the optical and near-infrared, and dust emissivity and the re-radiation of the light from young stars follows from measurements in the mid- and far-IR\null.  Unfortunately, 
this de-coupling is not complete, which renders the process of extracting information from a galaxy's SED a non-linear problem, with many local minima, co-variances between parameters, and highly non-Gaussian uncertainties.  (For extensive discussions of this topic, see \citealt{walcher+11} and \citealt{conroy13}.)  One therefore must employ a numerical procedure which can handle complex behavior for large numbers of parameters.  

One such method is the Markov-Chain Monte-Carlo (MCMC) technique, which is designed to explore all regions of a multi-dimensional parameter space while honing in on the highest likelihood parts of the probability distribution function. Over the past decade, MCMC algorithms have been employed for a myriad of problems in astronomy and astrophysics, including stellar population synthesis for galaxies and AGN, in both the near and distant universe \citep[i.e.,][]{GalMC, CIGALEMC, prospector}. 

Inspired by several recent works \citep[e.g.,][]{GalMC, prospector, bagpipes}, we introduce \mcsed , a modular stellar population synthesis code that takes libraries of simple stellar populations (SSPs) and mates a flexible method of creating composite populations with the MCMC fitting program, \texttt{emcee} \citep{emcee}.  The combination of these components provides an inference framework for the creation of models with parameters with well-characterized uncertainties for a galaxy's full UV through far-IR spectral energy distribution.  To demonstrate the capabilities of \mcsed , we apply the code to the sample of 1952 optical emission-line galaxies  identified by \citet{bowman+19} via the luminosity of their rest-frame optical emission lines. These $1.90 < z < 2.35$ systems have vigorous on-going star formation, with a young ($\lesssim 100$~Myr) population that far outshines the contribution of older ($\gtrsim$~Gyr) stars.  Moreover, because these galaxies were originally identified on {\sl HST\/} grism frames, they make an excellent test set for exploring the systematics of the types of galaxies that will be found by next-generation space-based grism surveys, such as those of planned for Euclid and WFIRST \citep{NISP, WFIRST}\null.  

Section \S\ref{sec:Sample} briefly describes our galaxy sample and the archival photometry and spectrophotometry available for analysis.  Section \S\ref{sec:SEDfit} lists the parameters used for modeling the galaxies' UV through near-IR spectral energy distributions and introduces our new SED fitting code, \mcsed. Section \S\ref{sec:results} explores the basic properties of the \citet{bowman+19} sample, the systematics of the dataset, and the effect that various assumptions have on the results.  Section \S\ref{sec:farIR} extends our analysis to the mid- and far-IR, and tests the ability of \mcsed\ to explore the balance between the sight-line attenuation of the UV stellar continua by dust and its isotropic re-radiation at longer wavelengths.  Finally, in \S\ref{sec:Discussion}, our results are placed in the context of other work at these redshifts and within the framework of simulations. 

Throughout this paper, we assume a $\Lambda$CDM cosmology, with $\Omega_{\Lambda} = 0.7$, $\Omega_M = 0.3$ and $H_0 = 70$~km~s$^{-1}$~Mpc$^{-1}$ \citep{bennett+13}.

\section{The Sample}
\label{sec:Sample}
At present, the largest samples of $z \gtrsim 2$ galaxies have been constructed from broadband photometry and color-selection via the use of photometric redshifts \citep[e.g.,][]{reddy+12, bouwens+15, ono+18}.  However, in the near future, space missions such as Euclid/NISP \citep{NISP} and WFIRST \citep{WFIRST} will detect millions of emission-line galaxies via near-IR slitless grism spectroscopy.  Since the selection criteria for these galaxies will differ substantially from those identified using broad-band photometric techniques, we can expect the distribution of the galaxies' physical properties to differ as well.

To prepare for this era, we can study the objects found by the pathfinding \threedhst\ survey, a 625~arcmin$^2$ set of 2-orbit observations taken with the Wide Field Camera 3 and the G141 grism on the {\sl Hubble Space Telescope\/}  \citep[GO programs 11600, 12177, and 12328;][]{3DHST, weiner+14}. This dataset consists of $R \sim 130$ near-IR ($1.08~\mu{\rm m}<\lambda < 1.68~\mu$m) slitless spectroscopy within the deep CANDELS fields of AEGIS, COSMOS, UDS, GOODS-N and GOODS-S \citep{AEGIS, COSMOS, UDS,  GOODS}.  Tens of thousands of objects are detectable on these frames, but of special interest are galaxies in the redshift range $1.90 < z < 2.35$.  In this window, the emission lines of \OII\ $\lambda 3727$, \NeIII\ $\lambda 3869$, H$\beta$, and the distinctively-shaped \OIII\ blended doublet $\lambda\lambda 4959,5007$ are simultaneously present in the bandpass.  This set of observable emission lines, which includes two different ionization states of oxygen, not only enables the determination of unambiguous redshifts, but also allows for the direct detection of galaxies over an extremely wide range of metallicity and ionization parameter.

\citet{bowman+19} recently created a sample of 1952 such optical emission line galaxies (oELGs) by vetting the list of $1.90 < z < 2.35$ objects in the \threedhst\ database \citep{3DHST}, and removing spurious detections, line mis-identifications, and known active-galactic nuclei. These systems, which generally have \OIII\ as their brightest feature, have sizes $R \lesssim 5$~kpc, \OIII\ $\lambda 5007$ fluxes brighter than $F \sim 4 \times 10^{-17}$~ergs~cm$^{-2}$~s$^{-1}$ (50\% completeness limit), and rest-frame near-IR magnitudes (IRAC 3 + IRAC 4) between $21.8 < m_{\rm JK} < 26.0$.  A full description of the demographics of this sample is given by \citet{bowman+19}.

\subsection{Photometry}
\label{sec:photom}
To define the galaxies' spectral energy distributions, we began with the SExtractor-based photometric catalog of \citet{skelton+14}.  This dataset starts with deep co-added $F125W + F140W + F160W$ images from {\sl HST\/} and then matches the data to the results of $\sim 20$ distinct ground- and space-based imaging programs. The result is a homogeneous, PSF-matched set of broad- and intermediate-band flux densities covering the observed wavelength range of $0.35~\mu$m to $8.0~\mu$m over the entire region surveyed by the {\sl HST\/} grism.  The poorest wavelength coverage is in the UDS field, which has photometry through 18 different bandpasses; the best dataset is in COSMOS, which has  imaging through 44 separate filters, including 12 intermediate band ($R \sim 20$) filters distributed between 4250~\AA\ and 8240~\AA\ \citep{skelton+14}.  For $z \sim 2$ systems, these data cover the rest-frame FUV through the rest-frame near-IR and form a homogeneous set of measurements for galaxies with $F140W$ magnitudes as faint as \mbox{$\sim 26$ mag}.

In addition to adopting the \citet{skelton+14} measurements, we examined the Rainbow Cosmological Surveys Database \citep{barro+11} for photometric data in the mid- and far-IR\null.  Formally, 93 of the \mbox{GOODS-S} and COSMOS galaxies in the \citet{bowman+19} sample have detections at $24~\mu$m from the {\sl Spitzer/}MIPS instrument, and 57 have data at longer wavelengths (via {\sl Spitzer/}MIPS and {\sl Herschel/}PACS observations).  However, the utility of many of these measurements is uncertain, due to their low signal-to-noise ratio (SNR) and possible confusion with superposed foreground sources.  Consequently, we do not use these data in our analysis, except for a proof-of-concept study described in \S\ref{sec:farIR}.

Finally, there is one additional constraint that can be applied to our SED fits: emission line fluxes from the \threedhst\ reduction of the WFC3 grism data \citep{3DHST}.  A significant fraction of the \citet{bowman+19} sample have H$\beta$ measurements (70\% with SNR $> 1$ and 25\% with SNR $> 3$), which can be used to constrain the galaxies' present-day star-formation rates \citep[e.g.,][]{kennicutt98, kennicutt+12}.  In addition, most of the systems have measurements (or limits) (70\% with SNR $> 1$ and 20\% with SNR $> 3$) on the excitation via the \OIII/H$\beta$ ratio \citep[e.g.,][]{kewley+01}.  \mcsed\ can use these fluxes to better constrain the state of the galaxies' star-forming populations. 

\section{Spectral Energy Distribution Modeling}
\label{sec:SEDfit}

SED fitting involves a considerable number of parameters.  Generally speaking, these can be divided into four categories: star formation history (SFH\null), stellar metallicity, dust attenuation, and dust emission. We describe them in detail below.

The first category involves the galaxies' SFH. Several studies have shown that for systems at $z \gtrsim 2$, constant or declining star formation histories are generally inappropriate \citep{reddy+12, pacifici+13, salmon+15}.  More realistic approaches model the SFH rate with three or four parameters, \citep[e.g.,][]{behroozi+13, simha+14}, use discrete star-formation rate (SFR) histories binned into physically-motivated age intervals \citep{prospector}, or adopt a dense basis approach, where non-monotonic and/or star-bursting behavior is reproduced using semi-analytic models that use cosmological simulations as a guide \citep{iyer+17}.  The assumptions made about the SFH will affect the estimates of a galaxy's stellar mass and age.

The second class of SED variable involves stellar metallicity. Broadband photometry generally provides only a weak constraint on the metal abundance(s) of a stellar population, as its effect is largely degenerate with those of other parameters, such as the system's SFH and dust attenuation \citep{worthey94, bell+01}.  However, if the data include high signal-to-noise ratio measurements of emission-line ratios and/or absorption-line indices, tighter constraints on the metallicity may be possible.  Thus, depending on the goal of an investigation, metallicity may be fixed, treated as a free parameter, or tied to another property, such as stellar mass \citep[e.g.,][]{peng+14, ma+16}.

The third type of SED variable involves the UV/optical attenuation by dust.   This issue is complicated: not only do different types of galaxies have different attenuation laws \citep[e.g.,][]{calzetti+00, conroy+10, wild+11}, but it is likely that the properties of the dust that affects young ($\lesssim 10^7$~yr) stellar populations differ from those which attenuates older stars \citep[e.g.,][]{cha-fall00}.  As a result, the choice of attenuation law propagates into estimates of other physical quantities, such as the recent star formation rate, stellar metallicity, and total $V$-band absorption.

The final category parameterizes the emission from warm and cold dust.  The energy budget of many star-forming galaxies includes a significant contribution from emission in the mid- and far-IR, so these components must be included in the SED fitting process.  Such models are complicated, however, as they need to include the behavior of graphites, olivine silicates, and PAH molecules as a function of photon, ion, and electron irradiation.  Nevertheless, several prescriptions for long-wavelength emission are available in the literature, including those of \citet{draine+07} and \citet{jones+17}.

Each of these parameter classes dominates the SED at a different set of wavelengths. Measurements in the rest-frame UV primarily carry information about dust absorption and recent star formation. In contrast, data in rest-frame optical and near-IR constrain a galaxy's star-formation history, metallicity, and stellar mass, while photometry in the mid- and far-IR reflect the emissivity of dust. It is due to this near de-coupling that SED fitting can be successful, even if complete spectral coverage is not available.  For example, if far-IR measurements do not exist, one can still use data in the rest-frame UV, optical, and near-IR to gain valuable insights into a galaxy's stellar populations and dust content.

\subsection{\mcsed }
\label{sec:mcsed}

As described above, a full UV through far-IR SED modeling of a galaxy may involve ten or more parameters.  One therefore requires an SED fitting program that is: (1) sufficiently general to handle a diverse range of inputs, (2) powerful enough to efficiently search through a multi-dimensional space and obtain realistic uncertainties on each parameter, and (3) flexible enough to allow the user to easily tailor the code to a specific problem or set of observational constraints. These conditions require the program to have many of the most commonly-used astrophysical relations built in, and be capable of accepting a wide variety of data, such as photometric measurements through broad- and intermediate-band filters, emission line fluxes from recombination and collisional-excitation, and absorption-line spectral indices.  Indeed, in the era of space missions such as Euclid and WFIRST, galaxies with both broadband photometry and emission-line spectrophotometry will be the rule, rather than the exception.  

\mcsed\ is constructed to handle dust-free spectra from a library of simple stellar populations.  The code creates a composite stellar population for a given star formation rate history, adds nebular emission, modifies the resulting spectrum using an assumed attenuation law, adds in mid- and far-IR emission from dust, and finally redshifts the composite spectrum to the observed distance. Since \mcsed\ is built to be modular, the user can easily change the basic fitting assumptions to suit their needs. Many of the most commonly used SED-fitting prescriptions (e.g., those pertaining to the SFH or dust attenuation) are already built into the program, and additional options, (such as alternative libraries for stellar and nebular emission) can be incorporated with relative ease.

The current version of \mcsed\ includes the library of SSPs from \texttt{FSPS}, the Flexible Stellar Population Synthesis code \citep{fsps-1, fsps-2, python-fsps}, and the prescription for nebular line and continuum emission given by the grid of \cloudy\ models \citep{CLOUDY, CLOUDY13} generated by \citet{byler+17}. One noteworthy feature of this combination of SSP and nebular models is their self-consistency: the prescription for nebular line and continuum emission, as a function of age, stellar metallicity, and ionization parameter, are based on the same SSP spectra that are used in FSPS\null.
In total, these SSPs consist of a grid of 22 metallicities ($-1.98 \leq \log Z/Z_{\odot} \leq +0.20$) and 84 ages ($6 \leq \log t({\rm yr}) \leq 10.15$), while nebular emission is modeled via a grid containing 11 metallicities ($-2.0 \leq \log Z/Z_{\odot} \leq +0.2$), 7 ionization parameters ($-4 < \log U < -1$), and 9 ages ($0.5 \leq t({\rm Myr}) < 10$).  

\mcsed\ includes several prescriptions for dust extinction and attenuation \citep[e.g.,][]{cardelli+89, calzetti+00, noll+09}. There are options for linking the attenuation of birth clouds and the diffuse dust component via a coefficient that can be set by the user (e.g.,  \mbox{$E(B-V)_{\rm diffuse} = 0.44 E(B-V)_{\rm birth}$}) or adopting separate attenuation laws for the two populations. A variety of options for star formation history are also provided in this release of \mcsed, including constant and exponentially increasing/decreasing SFHs, a double power-law SFH \citep{behroozi+13}, and a SFH defined via a set of user-defined age bins.

Dust emission is included using the parameterization of \citet{draine+07} and \citet{draine+07b}. This formulation is defined by the lower cutoff of the starlight intensity distribution ($U_{\rm min}$), the fraction of dust heated by starlight with $U > U_{\rm min}$ ($\gamma$), and the PAH mass fraction ($q_{\rm PAH}$).  The total dust mass ($M_{\rm dust}$) can either be treated as a free parameter to be fit via the normalization of the dust emission, or constrained via the energy balance of attenuation.  While the \citet{draine+07} models are based on Milky-Way dust, they have been successfully applied to high-redshift objects \citep{utomo+14}.

For a given set of SED parameters, \mcsed\ begins by constructing a set of composite stellar populations (CSPs) from a linear combination of SSPs defined by the star formation history.  Following \citet{mitchell+13}, \mcsed\ minimizes biases in stellar mass and other inferred quantities by avoiding the use of a single metallicity for this initial grid and instead introduces a small spread in abundance defined by a Gaussian kernel with dispersion $\sigma = 0.15 \log (Z / Z_{\odot})$ centered on the SED input metallicity.  After creating the CSP, \mcsed\ attenuates its spectrum, adds dust emission, redshifts the model to the observers' frame, and compares the result to the list of observational constraints, which can include photometry, emission-line fluxes, and absorption-line spectral indices.  Because some emission lines, such as those produced by collisional-excitation, are more difficult to model than others, the relative contribution of these lines to the overall likelihood can be adjusted by the user.

To perform the multi-dimensional parameter search, \mcsed\ uses the python package \texttt{emcee} \citep{emcee}, an MCMC algorithm which employs an affine-invariant ensemble sampler \citep{goodman+10}. Such an approach is insensitive to co-variances in the fitted parameters, and is thus well-suited for efficiently exploring the high-dimensionality and oddly-shaped likelihood distributions that often occur in SED fitting.  In addition, \texttt{emcee} generally requires no manual input or running period to tune the proposal distributions, as is common in Metropolis-Hastings MCMC algorithms. As \texttt{emcee} is written in \texttt{python}, it nicely fits into the collection of \texttt{python} packages used here.

As describe above, the many run-time options built into \mcsed\ allow the user to select from a wide variety of SFHs and dust attenuation curves, or implement their own parameterizations for these variables.  Configuration options are also available for the inclusion of new photometric filters and SSP models. Finally, \mcsed\ features a test mode, where the user can create and ``observe'' model galaxies and compare their inferred parameters to the input ``truth''.  Details about \mcsed\ are given in Appendix~\ref{appendix:a}\null.  Appendix~\ref{appendix:b} uses \mcsed's test mode to evaluate the precision of recovering input parameters.  

\begin{figure*}[ht!]
  \captionsetup[subfigure]{labelformat=parens}
  \centering
  \noindent\includegraphics[width=\linewidth]{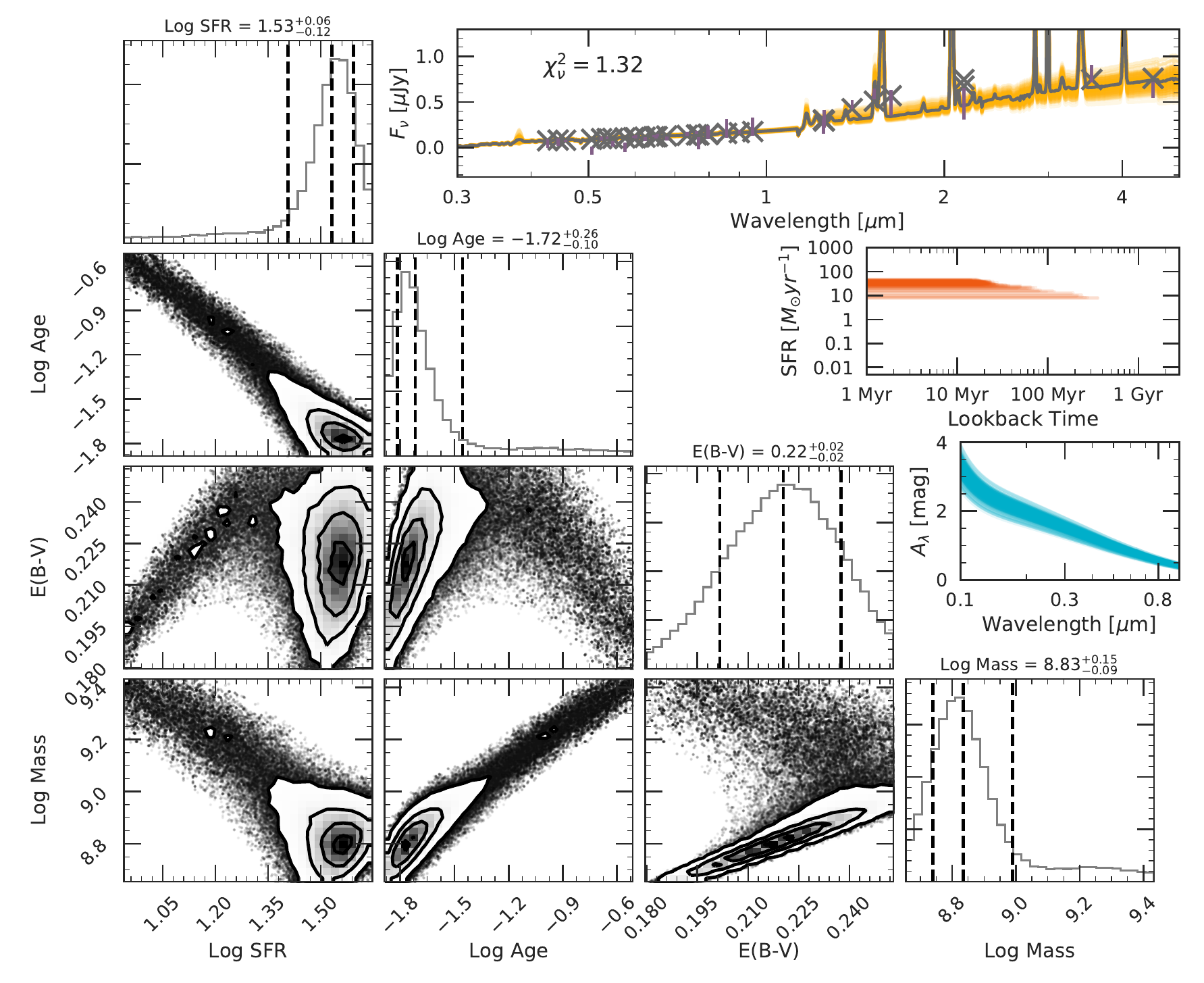}
  \caption{Diagnostic figure for a $z = 2.14$ emission-line galaxy (GOODS-S~43481), including the parameter co-variances and marginalized probability distributions for the history of star formation, the dust attenuation curve, and best-fit SED\null. This example adopts a simplified set of fitting assumptions, including a constant SFH, a \citet{calzetti+00} attenuation law, and a fixed stellar metallicity of 40\% solar.  The rightmost panel of each row displays the likelihood histogram for the captioned variable, with the most likely value (and its $1 \, \sigma$ confidence interval) denoted by dashed lines.    The three panels in the top right corner show  randomly-drawn realizations of the posterior distributions.  
  }
  \label{fig:triangle_simple}
\end{figure*}

An example of \mcsed's output is shown in Figure~\ref{fig:triangle_simple}.  Displayed are the probability distributions for a typical $z \sim 2$ emission-line galaxy, along with co-variance plots, fitted SFR histories and attenuation curves, best-fit SEDs, and modeled and observed filter fluxes and emission-line strengths.  The example shown uses a simplified set of fitting assumptions, i.e., a constant SFR, a \cite{calzetti+00} dust attenuation law, a fixed stellar metallicity of $Z = 0.0077$ ($40\%$ solar), a fixed nebular ionization parameter of $\log U = -2$, and a fixed \citet{draine+07} dust-emission model with $U_{\rm min}=2.0$, $\gamma=0.05$, and $q_{\rm PAH}=2.5$.  \mcsed\  can easily be reconfigured to accommodate a more realistic set of fitting assumptions.  In addition to the diagnostic plot shown in Figure~\ref{fig:triangle_simple}, many other \mcsed\ outputs are available on demand, including a summary table of the best-fit model parameters and (user-defined) confidence intervals, the full posterior probability distributions of the model parameters, modeled and observed filter and emission-line fluxes, the best-fit SED, and a log file detailing the full set of parameters and fitting assumptions that were adopted for the run. 

\section{Fitting Emission-Line Galaxies from \threedhst}
\label{sec:results}
As described in \S\ref{sec:photom}, all the galaxies in the five CANDELS fields have comprehensive multi-wavelength photometry which extends from the atmospheric cutoff near 3500~\AA\  to the {\sl Spitzer\/}/IRAC 3.6, 4.5, 5.8, and $7.9~\mu$m bands \citep{skelton+14}. In the \citet{bowman+19} redshift window of $1.9 < z < 2.35$, there are also measured fluxes (or limits) for the strong emission lines within the spectral range $\sim 3700$ to $\sim 5100$~\AA\ from the WFC G141-grism spectroscopy of the \threedhst\ team \citep{3DHST}. These grism data provide an important constraint for fitting the galaxies' SEDs, as the relative strengths of \OIII\ $\lambda\lambda 4959,5007$, \OII\ $\lambda 3727$, and H$\beta$ reflect the metallicity of the galaxy's ISM and, implicitly, the metallicity of the latest generation of stars.  At the same time, the absolute strengths of these emission lines indicate the amount of star formation currently taking place in the galaxy ($t < 10^7$~yr).  What the \citet{bowman+19} sample lacks is information in the mid- and far-IR:  only $\sim 15\%$ of the galaxies have measurements with the {\sl Spitzer\/}/MIPS detector at $24~\mu$m (rest-frame wavelength near $8~\mu$m) and just 143 objects ($7\%$) have data at longer wavelengths.  Thus, we start our analysis by reducing the dimensionality of the problem by fixing the dust emissivity variables to $U_{\rm min}=2.0$, $\gamma=0.05$, $q_{\rm PAH}=2.5$, and $M_{\rm dust} = 10^7 M_\odot$.  We will return to this point when we examine the SEDs of two galaxies that have reliable longer-wavelength data.

\subsection{Fitting Assumptions}
\label{sec:fitting-assumptions}

As discussed by \citet{walcher+11}, \citet{conroy13} and references therein, the physical properties one obtains from SED fitting depend in a non-trivial manner on the fitting assumptions one uses in the analysis. Moreover, these trends may change from population to population, as the systematics one obtains for relatively quiescent galaxies may be substantially different from those of starburst galaxies.  Because of this behavior, it is useful to explore the effect that each of our underlying assumptions has on the derived properties of moderate redshift ($z \sim 2$) emission-line galaxies, as many millions of such objects will be identified by upcoming missions such as Euclid and WFIRST\null.

\begin{figure}[ht!]
  \captionsetup[subfigure]{labelformat=parens}
  \centering
  \noindent\includegraphics[width=\linewidth]{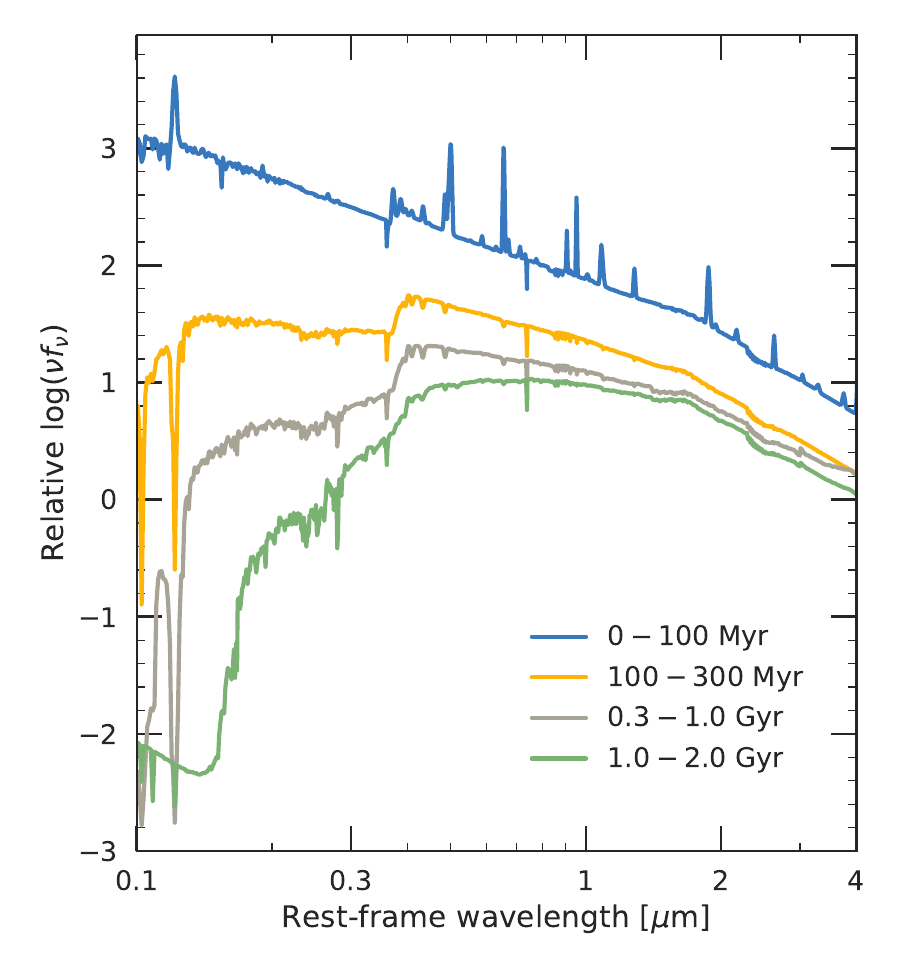}
  \caption{ {Sample SEDs for the four age-bins in our SFH\null.  Since our galaxies are all at $z \geq 1.9$ the SSP ages need only extend 2~Gyr. In this example, the stellar metallicity has been fixed at \mbox{$Z = 0.0077$} and the nebular emission assumes an ionization parameter of \mbox{$\log U = -2$}.}
  }
  \label{fig:direct_ssp}
\end{figure}

While many studies have demonstrated that simplified fitting assumptions (e.g., a constant SFH, a fixed stellar metallicity, and a \citet{calzetti+00} attenuation law) over-constrain galaxy properties across a range of redshifts and galaxy types \citep[e.g.,][]{reddy+12, pacifici+13, salmon+15}, the ``optimal'' set of fitting assumptions for $z \sim 2$ emission-line galaxies remains an open question.  Fortunately, the wealth of data available in the CANDELS fields allows a relaxation of these assumptions.   Our baseline analysis begins by dividing the galaxies' star-formation rate histories into four bins with log ages (in years) between [$6.0 - 8.0$], [$8.0 - 8.5$], [$8.5 - 9.0$] and [$9.0 - 9.3$], with a constant SFR within each bin. As noted by \citet{conroy13} and \citet{prospector}, fits with parameterized forms of the SFH may be susceptible to poorly-quantified systematic uncertainties. We adopt the \citet{noll+09} formulation for dust attenuation internal to the galaxies, allow stellar metallicity to be a free parameter, and, motivated by the high \OIII/[\ion{O}{2}] and \OIII/\Hb\ ratios present in the \citet{bowman+19} sample, set the ambient ionization parameter to $\log U = -2$.  This procedure leaves eight unknowns:  one for each age bin of the SFH, three for attenuation (the slope of the UV wavelength dependence, the strength of the 2175~\AA\ bump, and the total amount of extinction, which we parameterize via $E(B-V)$), and one for stellar metallicity.  We adopt a coefficient ($0.44$~mag; \citealt{calzetti+00}) to link the total attenuation of young stars still enshrouded in their birth clouds ($t\leq 10^7$~yr) to that affecting older stars.  A list of the eight free parameters and their priors is given in Table~\ref{tab:parameters}, along with several values that are held fixed throughout the fitting.  Sample SSP spectra, which contribute to each age bin in our SFH, are presented in Figure~\ref{fig:direct_ssp}. 

\begin{deluxetable*}{lllr}
\tablecaption{Free and Fixed Model Parameters
\label{tab:parameters}}
\tablehead{
\colhead{Parameter}
&\colhead{Description}
&\multicolumn{2}{c}{Priors}
}
\startdata
\multicolumn{3}{c}{Star Formation History} \\
sfr$_1$  & log(SFR) [M$_\odot$/yr] between $6.0 \leq \log t {\rm (yr)}  < 8.0$  & [$-5$, 3]  &(uniform) \\
sfr$_2$  & log(SFR) [M$_\odot$/yr] between $8.0 \leq \log t {\rm (yr)}  < 8.5$  & [$-5$, 3]  &(uniform) \\
sfr$_3$  & log(SFR) [M$_\odot$/yr] between $8.5 \leq \log t {\rm (yr)}  < 9.0$  & [$-5$, 3]  &(uniform) \\
sfr$_4$  & log(SFR) [M$_\odot$/yr] between $9.0 \leq \log t {\rm (yr)}  < 9.3$  & [$-5$, 3] &(uniform) \\ \\
\multicolumn{3}{c}{Attenuation Curve \citep{noll+09}}\\
$\delta$  &UV Slope  & [$-1$, 1]  &(uniform) \\
$E_b$    &Depth of 2175~\AA\ Feature  & [$-0.2$, 6.0]  &(uniform) \\
$E(B-V)$  &Total Attenuation & [$-0.05$, 1.50]  &(uniform) \\
$E(B-V)_{\rm diffuse}$ & Attenuation of the diffuse component (relative to the birth cloud) & 0.44 &(fixed) \\
$t_{\rm birth}$ & Age separating the birth cloud and diffuse components & 10 Myr &(fixed) \\ \\
\multicolumn{3}{c}{Stellar Population} \\
$\log(Z_{\rm stars} / Z_\odot)$       &Metallicity & [$-1.98$, 0.20] &(uniform) \\ \\
\multicolumn{3}{c}{Dust Emissivity \citep{draine+07}} \\
$U_{\rm min}$ &Lower cutoff of the starlight intensity distribution & 2.0 &(fixed) \\
$\gamma$      &Fraction of dust heated by starlight with $U > U_{\rm min}$  & 0.05 &(fixed) \\
$q_{\rm PAH}$ &PAH mass fraction & 2.5 &(fixed) \\ \\
\multicolumn{3}{c}{Nebular Emission} \\
$\log U$      &Log Ionization Parameter &$-2$ &(fixed) \\
$Z_{\rm gas}$    &Gas Phase Abundance &Set to $Z_{\rm stars}$ &\dots \\
$w(\lambda 3727)$ &Weight of \OII\ $\lambda \lambda 3727, 3729$ doublet & 0.5 &(fixed) \\
$w({\rm H}\beta$) &Weight of H$\beta$ line flux  &1.0 &(fixed) \\
$w(\lambda 5007)$ &Weight of \OIII\ $\lambda 5007$ line &0.5 &(fixed) \\
\enddata
\end{deluxetable*}

\subsection{The Effects of Assumptions on \mcsed\ Results}
\label{subsec:assumptions}

\begin{figure*}[hp!]
  \captionsetup[subfigure]{labelformat=empty}
  \centering
  \subfloat[][]{\label{fig:va-a}%
    \includegraphics[height=0.24\textheight]{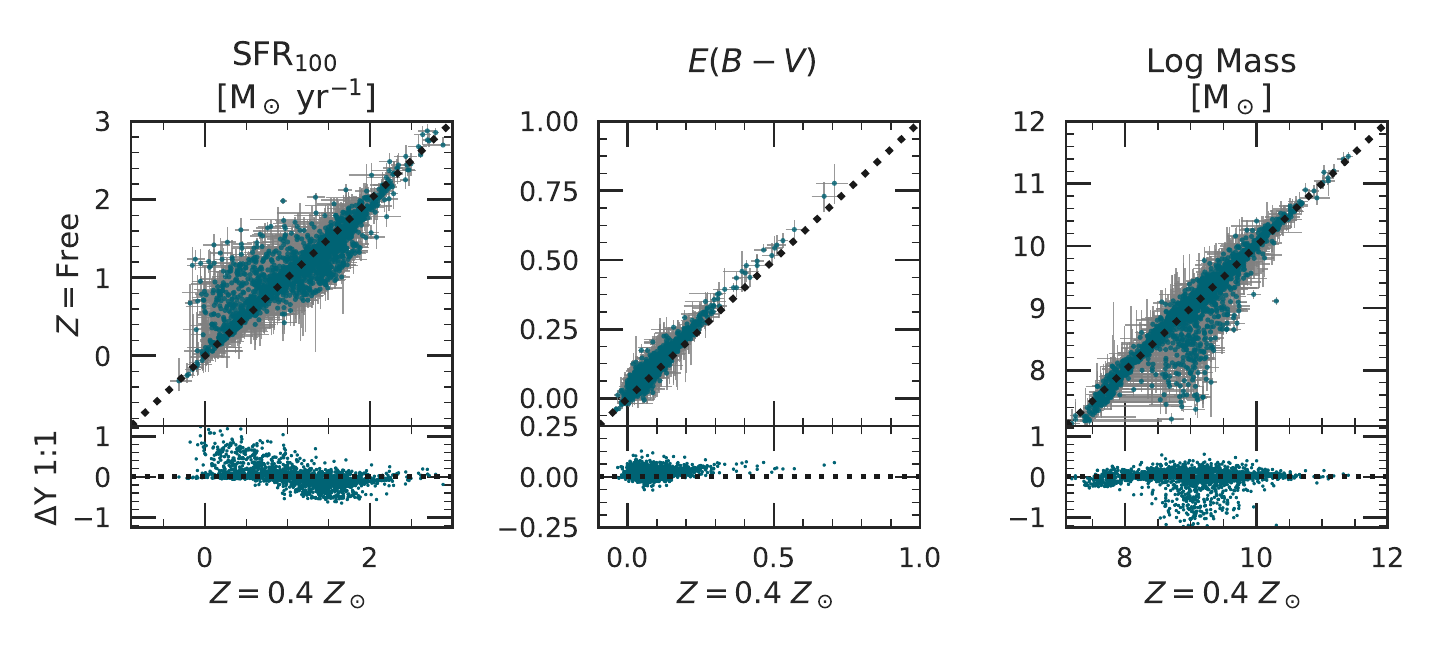}%
  }
  \\[-6ex]
  \subfloat[][]{\label{fig:va-b}%
    \includegraphics[height=0.24\textheight]{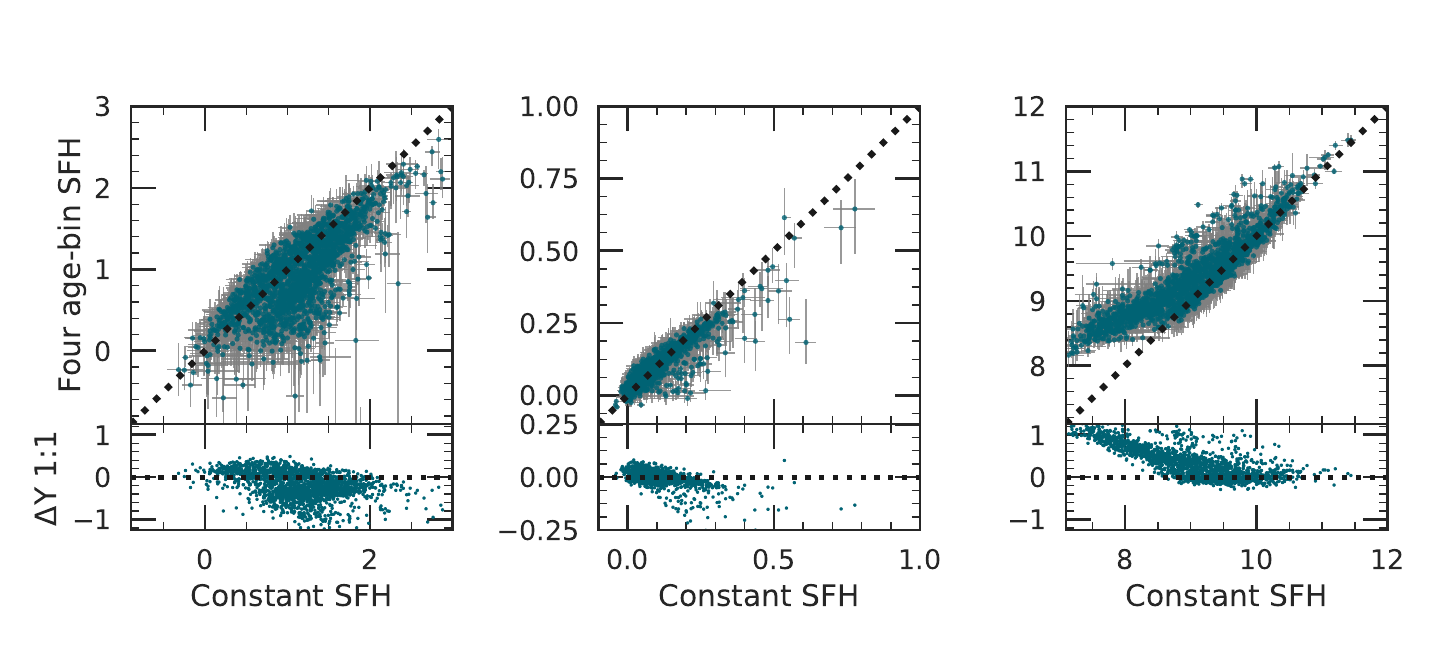}%
  }
  \\[-6ex]
  \subfloat[][]{\label{fig:va-c}%
    \includegraphics[height=0.24\textheight]{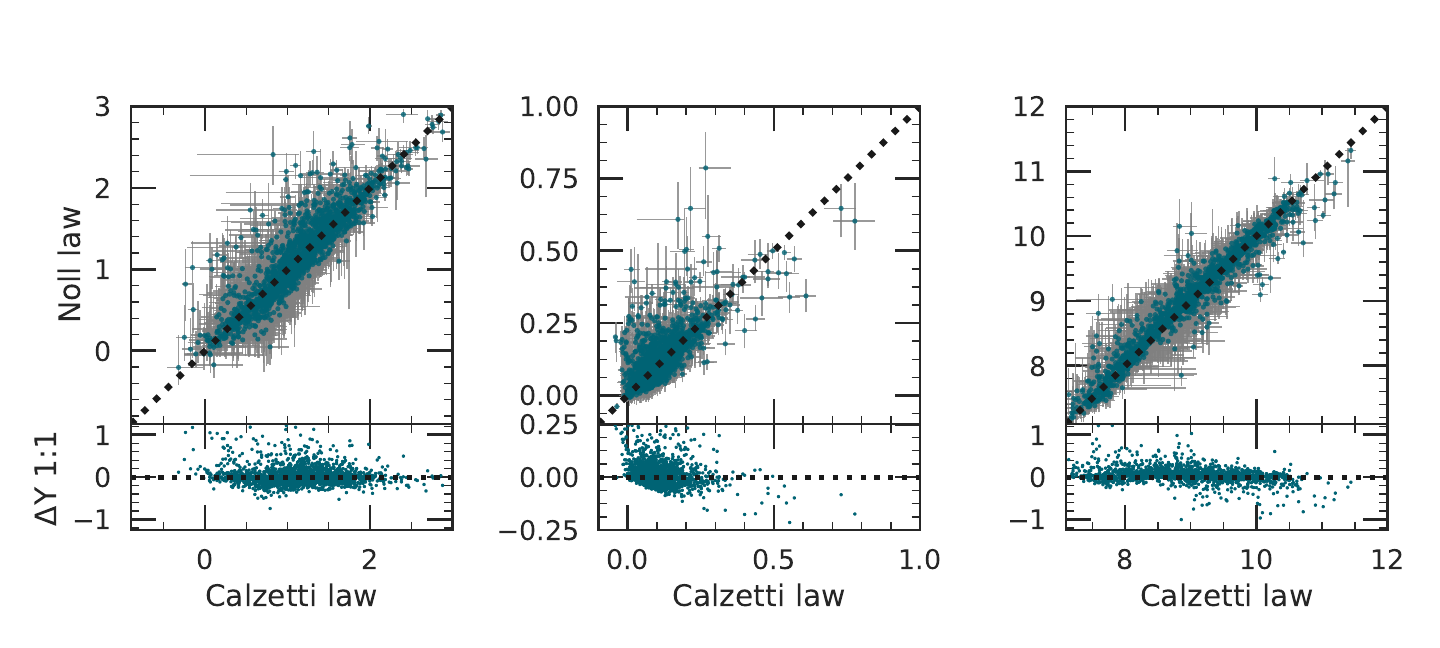}%
  }
  \caption{\mcsed\ solutions for (from left to right) the current star-formation rate, differential reddening, and stellar mass for the \citet{bowman+19} sample of $z \sim 2$ emission-line galaxies.  {\it Top row:} A comparison of the results for fixed metallicity ($x$-axis) with measurements where the system metallicity is left as a free parameter, while assuming a constant SFH and a \citet{calzetti+00} attenuation law.  {\it Middle row:} A comparison of the results for a constant SFR ($x$-axis) with the four age-bin SFH ($y$-axis), while assuming a \citet{calzetti+00} attenuation law and leaving stellar metallicity as a free parameter.  {\it Bottom row:}  A comparison of the results based on the \citet{calzetti+00} attenuation law ($x$-axis) with those based on the \citet{noll+09} law ($y$-axis), while adopting a constant SFH and leaving stellar metallicity as a free parameter.  
  }
  \label{fig:varying-assumptions}
\end{figure*}

\begin{figure*}[ht!]
  \captionsetup[subfigure]{labelformat=parens}
 \centering
  \subfloat[][]{\label{fig:fchange_EBV}%
    \includegraphics[width=0.5\linewidth]{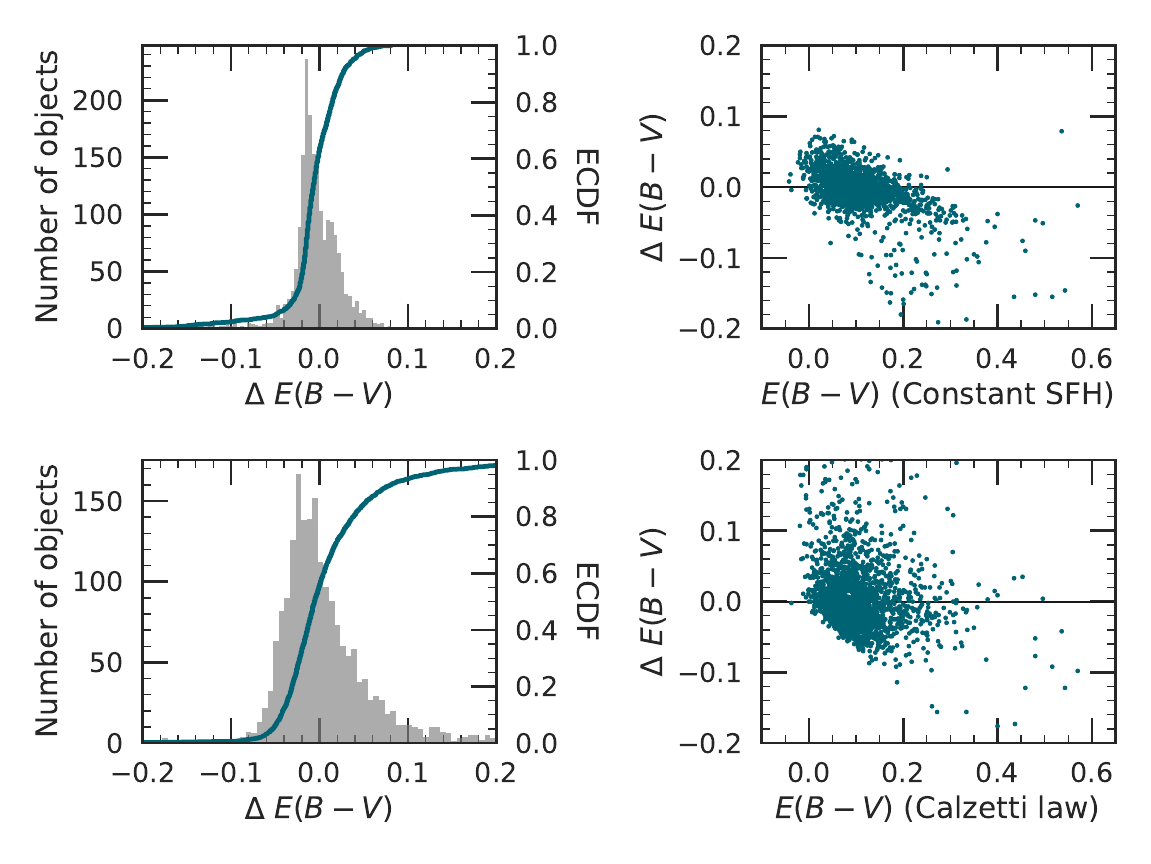}%
  }\hfill
  \subfloat[][]{\label{fig:fchange_SFR}%
    \includegraphics[width=0.5\linewidth]{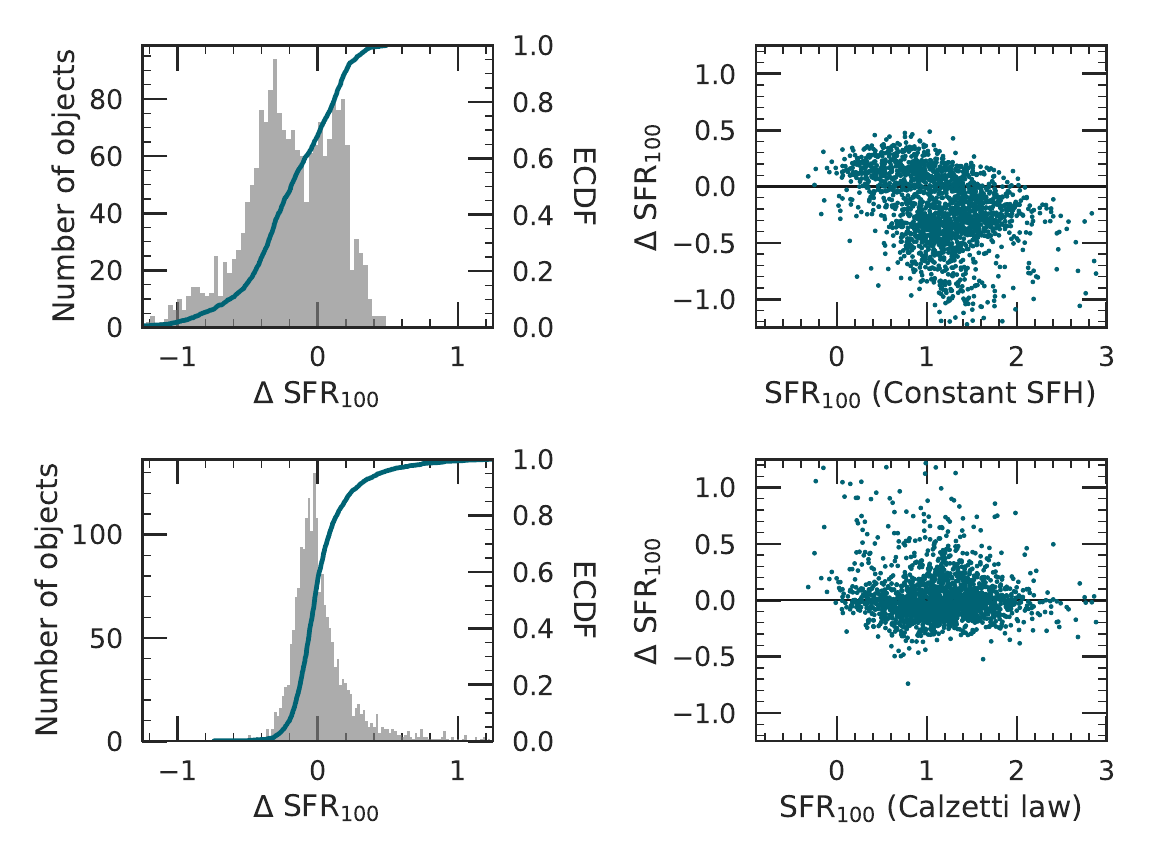}%
  }
  \caption{{\it Panel~(a), top row:} The difference between $E(B-V)$ values computed using the assumption of a constant SFR history and those found using a SFH divided into four age-bins.   These values assume a \citet{calzetti+00} dust law and were calculated using metallicity as a free parameter, but the results are similar for fixed metallicity or the \citet{noll+09} prescription for attenuation.  The red line on the left panel gives the cumulative distribution function.  {\it Panel~(a), bottom row:}  The difference between $E(B-V)$ values computed using the \citet{calzetti+00} attenuation law and those found using the \citet{noll+09} law.  Again, the values shown assume a constant SFH and have metallicity as a free parameter. As above, the results are largely unaffected by these latter two assumptions.  
  {\it Panel~(b), top row:} The difference in present day ($t < 100$~Myr) SFRs computed assuming a constant SFR history and those found using a SFH divided into four age-bins.  The values shown assume a \citet{calzetti+00} dust law and were calculated with metallicity as a free parameter.  The solid red line in the left panel indicates the cumulative distribution function.  {\it Panel~(b), bottom row:} The difference in present day SFRs computed using the \citet{calzetti+00} attenuation law and those found from the \citet{noll+09} law. The values assume a constant SFH and again have metallicity as a free parameter.
  }
  \label{fig:fchange_EBV_SFR}
\end{figure*}

Before analyzing the global properties of our galaxy sample, we investigate how the parameterizations listed above affect the best-fit solutions found by \mcsed\null.  This issue can be examined by modifying our assumptions, and, one by one, seeing what effect each has on the inferred properties of the galaxies.  The results of this exercise are shown in Figure~\ref{fig:varying-assumptions}, where three fundamental outputs of \mcsed\ are compared: the current star-formation rate, the total dust attenuation, as parameterized by $E(B-V)$, and the total stellar mass. The three rows of panels each vary one fitting assumption (stellar metallicity, SFH, and dust attenuation law in the top, middle, and bottom rows, respectively) while holding the other two assumptions fixed. The first row of panels compares parameter estimates inferred for fits with fixed metallicity at $Z = 0.0077$ ($x$-axis) to those where metallicity is kept as a free parameter ($y$-axis), while assuming a constant SFH and a \citet{calzetti+00} attenuation law. In the middle row, where metallicity is left as a free parameter, we adopt a \cite{calzetti+00} attenuation law and compare the results for a constant SFR history ($x$-axis) with the four age-bin SFH ($y$-axis). Finally, the bottom row of panels leaves metallicity as a free parameter and assumes a constant SFH, and compares the results from the \citet{calzetti+00} attenuation law ($x$-axis) to those from the more general \citet{noll+09} formulation ($y$-axis).
Histograms and differential diagrams of these comparisons are presented in Figure~\ref{fig:fchange_EBV} and \ref{fig:fchange_SFR}.

Figure~\ref{fig:varying-assumptions} displays how the inferred star-formation rate, dust attenuation, and stellar mass depend upon the SED fitting assumptions. While there are systematics and biases between the parameter estimates (discussed in more detail below), an examination of the figure demonstrates that the properties derived by \mcsed\ for our sample of vigorously star-forming galaxies are, broadly speaking, consistent across the various assumptions about star formation history, attenuation law, and stellar metallicity. 
The exact value of the best-fit recent SFR does depend on the systems' assumed SFR history (Figure~\ref{fig:varying-assumptions}, middle row) and (to a lesser extent) metallicity (Figure~\ref{fig:varying-assumptions}, top row), but these systematics are generally small. Interestingly, the choice of attenuation law does not cause a significant systematic shift in the SFR estimates, and the internal uncertainties in the measurements, as estimated by a series of re-sampling experiments with each point perturbed by its measurement error, is consistent with the panel's observed scatter about the 1:1 relation at the 90\% confidence level.

Figure~\ref{fig:varying-assumptions} also demonstrates that, while our assumptions about stellar metallicity have little effect on our conclusions, the best-fit $E(B-V)$ value does depend upon what one chooses for the star formation rate history and attenuation law (middle and bottom rows of Figure~\ref{fig:varying-assumptions}).  This systematic is {\it not} due to incorrect fitting by \mcsed\ (see  Appendix~\ref{appendix:b}), and is expected, as the \citet{noll+09} expression contains a parameter, $\delta$, that changes the wavelength dependence of the relation. For $z \sim 2$ galaxies, the bulk of the photometric measurements are in the rest-frame ultraviolet, so the added flexibility provided by $\delta$ propagates into a change in the optical reddening.  The dependence of $E(B-V)$ on SFH is less straightforward, but still apparent: more realistic SFHs with four age-bins produce slightly lower values of $E(B-V)$, independent of which attenuation law is used.  For $\sim 90\%$ of the galaxies, however, the reddening parameter is consistent to $\Delta E(B-V) < 0.1$ (Figure~\ref{fig:fchange_EBV}).

Perhaps the most interesting dependence shown in Figure~\ref{fig:varying-assumptions} is that for stellar mass.  Not surprisingly, the stellar mass estimates have little dependence on attenuation (Figure~\ref{fig:varying-assumptions}, bottom row), as the former is mostly derived from the near-IR while the latter's effect is predominantly in the ultraviolet.  Similarly, system metallicity has a relatively minor effect on stellar mass, especially for systems with $M \gtrsim 10^9 M_{\odot}$, although there are obvious systematics for a subset of the objects (Figure~\ref{fig:varying-assumptions}, top row). However, at lower masses, the assumption of a constant SFR produces mass estimates that are up to $\sim 1$ dex lower than those derived using the four age-bin SFH.  Clearly, the oversimplified assumption of a constant SFH can result in strong systematic shifts in low mass galaxies. A similar result appears when using a constant SFH and varying the systems' ionization parameter.

\begin{figure*}[ht!]
  \captionsetup[subfigure]{labelformat=parens}
  \centering
  \noindent\includegraphics[width=\linewidth]{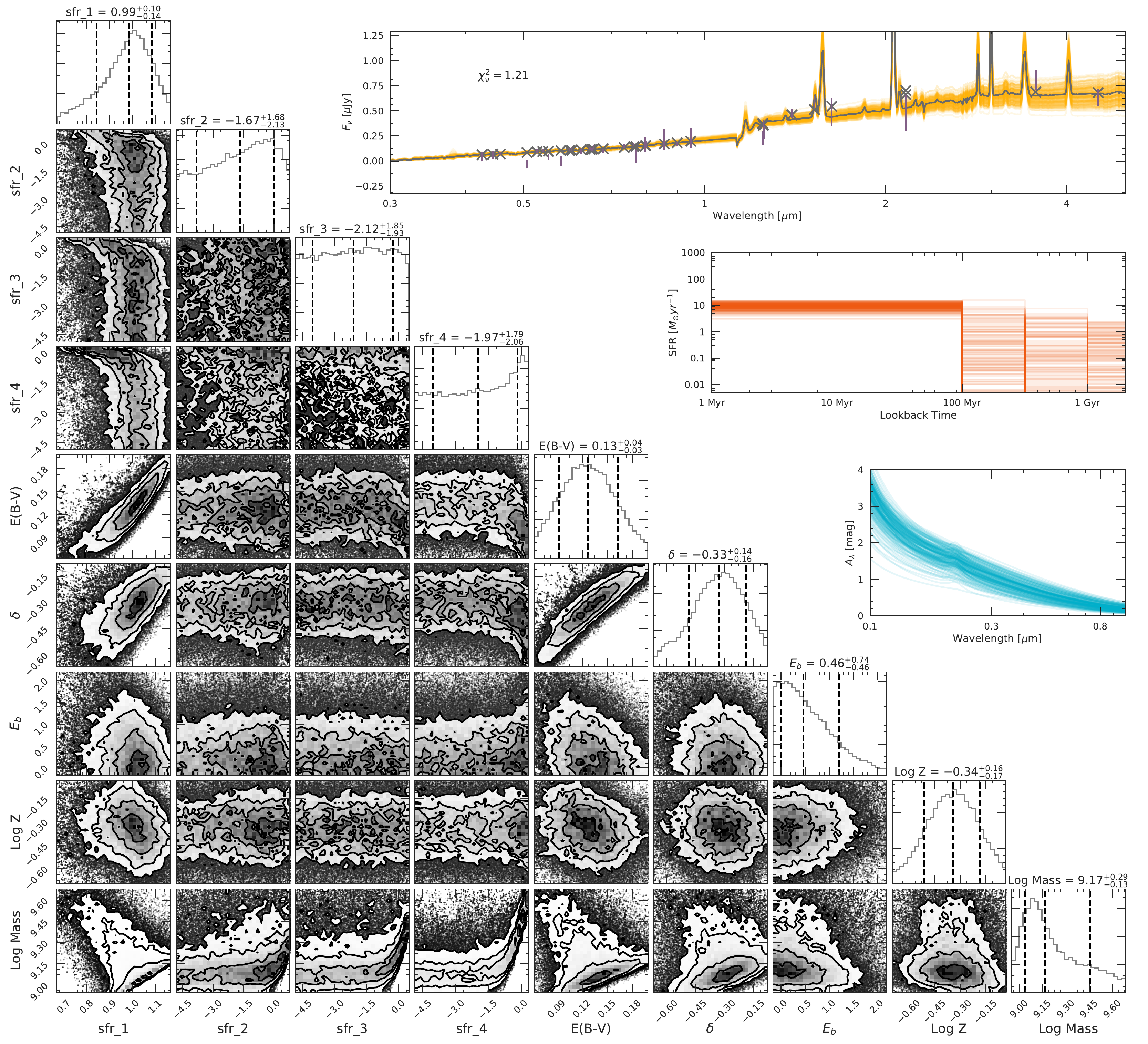}
  \caption{Diagnostic figure for the GOODS-S field galaxy shown in  Figure~\ref{fig:triangle_simple} using our flexible, eight-parameter model (four SFR age-bins, the \citet{noll+09} attenuation law, and metallicity as a free parameter).  As in Figure~\ref{fig:triangle_simple}, the curves in the top right three panels are randomly-drawn realizations of the posterior distributions, and the rightmost panel of each row gives a histogram of the likelihoods.  Despite the wealth of observational data available for the galaxy, parameters such as the 2175~\AA\ dust bump and the SFR at lookback times beyond a few Myrs are difficult to constrain.}
  \label{fig:triangle}
\end{figure*}

Both sets of fitting assumptions yield similar quality of fits to the data (with similar $\chi^2$ values). However, the diagnostic summary for one of the galaxies shown in Figure~\ref{fig:triangle} illustrates the unavoidable difficulty associated with estimating some aspects of $z\sim2$ optical emission-line galaxies. In this example, the recent SFR ($t < 100$~Myr), $E(B-V)$, and the stellar mass are fairly well constrained, as they are closely linked to the available data. (The emission line fluxes and the wealth of rest-frame UV photometry strongly constrain the recent SFR and $E(B-V)$, while the rich dataset at rest-frame wavelengths $\lambda \gtrsim 6000$~\AA\ define the stellar mass.) Other properties, such as the distant-past SFR and the strength of the 2175~\AA\ extinction bump, are, at best, weakly constrained.

\subsection{The Physical Properties of oELGs}

\begin{figure*}[ht!]
  \captionsetup[subfigure]{labelformat=parens}
  \centering
  \subfloat[][]{\label{fig:stacked-sed1}%
    \includegraphics[width=0.5\linewidth]{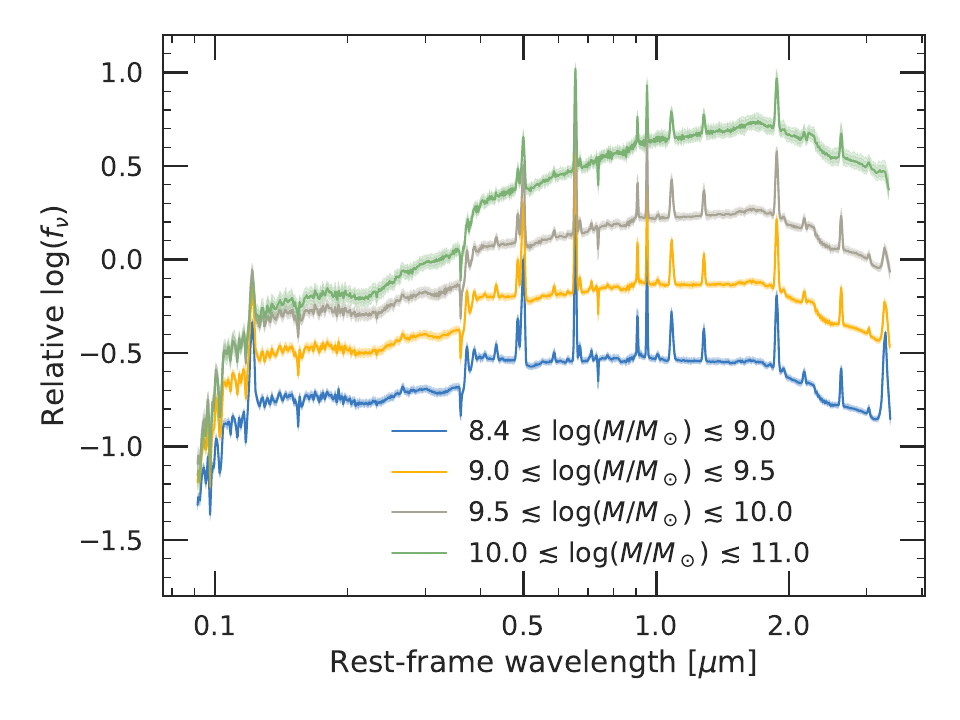}%
  }\hfill
  \subfloat[][]{\label{fig:stacked-dust1}%
    \includegraphics[width=0.5\linewidth]{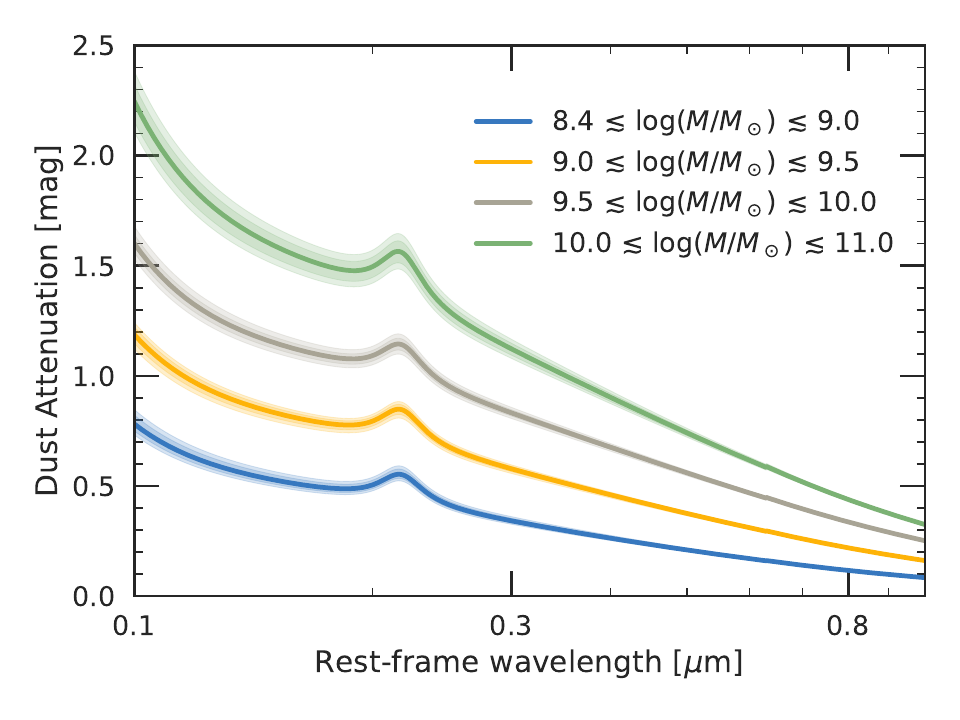}%
  }
  \\
  \subfloat[][]{\label{fig:stacked-sed2-dereddened}%
    \includegraphics[width=0.5\linewidth]{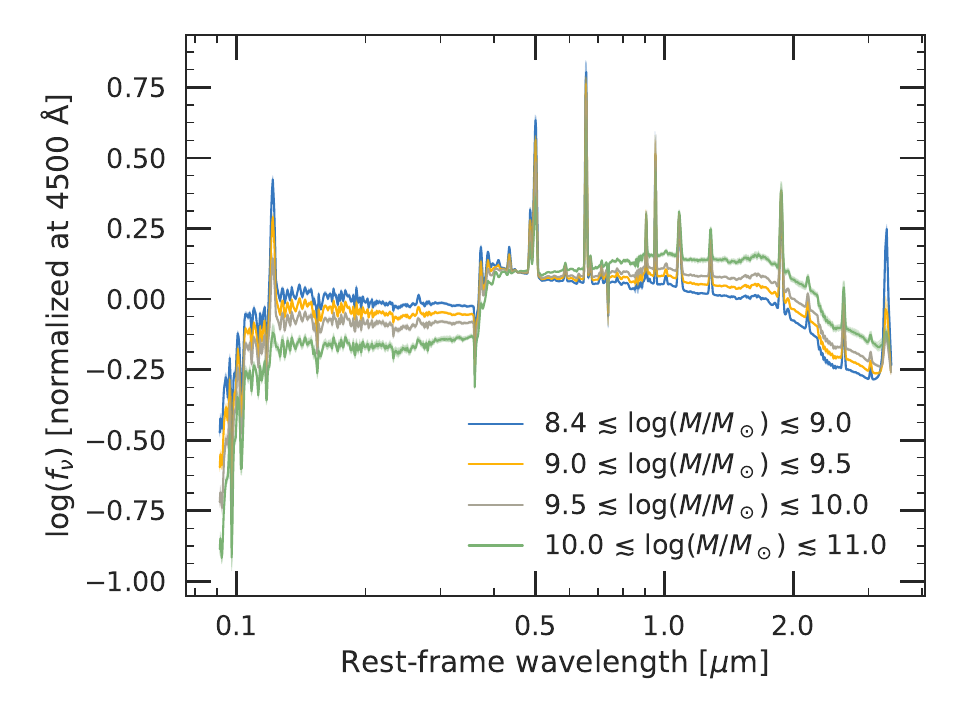}%
  }\hfill  
  \subfloat[][]{\label{fig:stacked-sfh2}%
    \includegraphics[width=0.5\linewidth]{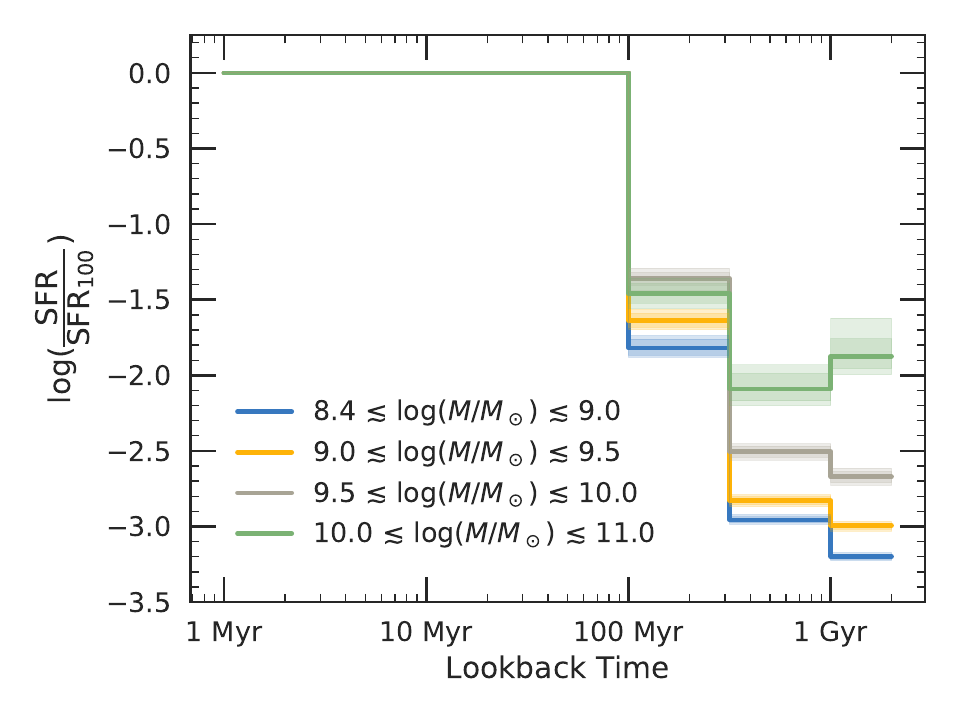}%
  }
  \caption{The average spectrum, dust attenuation curve, and star-formation rate history of $z\sim2$ emission-line galaxies spanning nearly three orders of magnitude in stellar mass.  As stellar mass increases, the energy budget shifts towards longer wavelengths, due to an increase in the overall dust content and a higher distant-past SFR. 
  }
  \label{fig:stacked-plots}
\end{figure*}

Our flexible fitting assumptions are next used to explore the variation of the dust attenuation curve, star formation rate history, and the behavior of the SED across nearly three orders of magnitude in stellar mass.  We divide our galaxy sample into four stellar mass-bins (each containing $\gtrsim 200$ objects), with Bin 1 having $\log M/M_{\odot} \leq 9.0$, Bin 2 with $9.0 < \log M/M_{\odot} \leq 9.5$, Bin 3 with $9.5 < \log M/M_{\odot} < 10.0$, and Bin 4 with $\log M/M_{\odot} \geq 10.0$.
For each bin, a representative spectrum is computed by finding the best-fitting SED of each galaxy, and taking the mean of the distribution.  Similarly, we calculated the dust-corrected, normalized SED of each bin by dereddening each galaxy (using its best-fitting attenuation parameters), normalizing its spectrum to the flux at 4500~\AA, and taking the bi-weight of the distribution.  Finally, to compute representative attenuation curves and SFHs, 
we stacked all the parameters estimates from each realization of the posterior distribution of each galaxy in the bin and computed the bi-weight of the distribution.
The $1 \, \sigma$ uncertainties (and 90\% confidence intervals) were found by bootstrapping these data.  Figure~\ref{fig:stacked-plots} displays the results of this analysis.

It is obvious from Panel~\ref{fig:stacked-sed1} that the SEDs of $z \sim 2$ emission-line galaxies vary systematically with mass: as the stellar mass increases, the galactic emission is shifted towards longer wavelengths. This unsurprising result can be explained by two effects: an increase in the amount of attenuation due to dust and a rise in the contribution of older, redder stars. The former effect is demonstrated in Panel~\ref{fig:stacked-dust1}, which displays the dust attenuation curves computed from the bi-weight of the \citet{noll+09} parameters.  This figure reveals that galaxies with stellar masses greater than $\sim 10^{10} M_{\odot}$ have a magnitude more near-UV attenuation than lower-mass ($M \lesssim  10^9 \, M_{\odot}$) systems. Dereddening the best-fit spectra prior to stacking (using the best-fit attenuation law parameters for each galaxy) produces Panel~\ref{fig:stacked-sed2-dereddened}, which presents the dust-free and flux-normalized version of the stacked SEDs.  This panel illustrates the different mix of stellar populations, with higher mass galaxies having a relatively larger amount of flux in the near-IR\null. Panel~\ref{fig:stacked-sfh2} confirms this result by showing the SFRs inferred for each of the four age-bins:  the highest stellar mass galaxies have relatively larger SFRs at older lookback times.

Beyond the increase in total dust content with higher stellar mass, several studies have shown that the wavelength-dependence of attenuation can vary from galaxy to galaxy. We address this through of use of the \citet{noll+09} law, which contains three parameters: $\delta$, which is the greyness/tilt of the wavelength dependence of attenuation relative to the \citet{calzetti+00} law, $E_b$, which represents the $\sim 350$~\AA\ FWHM ``bump'' of excess attenuation at 2175~\AA, and the total amount of extinction, parameterized by $E(B-V)$.  The mass-dependent trends shown in Figure~\ref{fig:stacked-dust1} are primarily due to an increase in $E(B-V)$, but we can also use our SED analysis to test for variations in $\delta$ and $E_b$.

\begin{figure}[h!]
  \captionsetup[subfigure]{labelformat=parens}
  \centering
  \noindent\includegraphics[width=\linewidth]{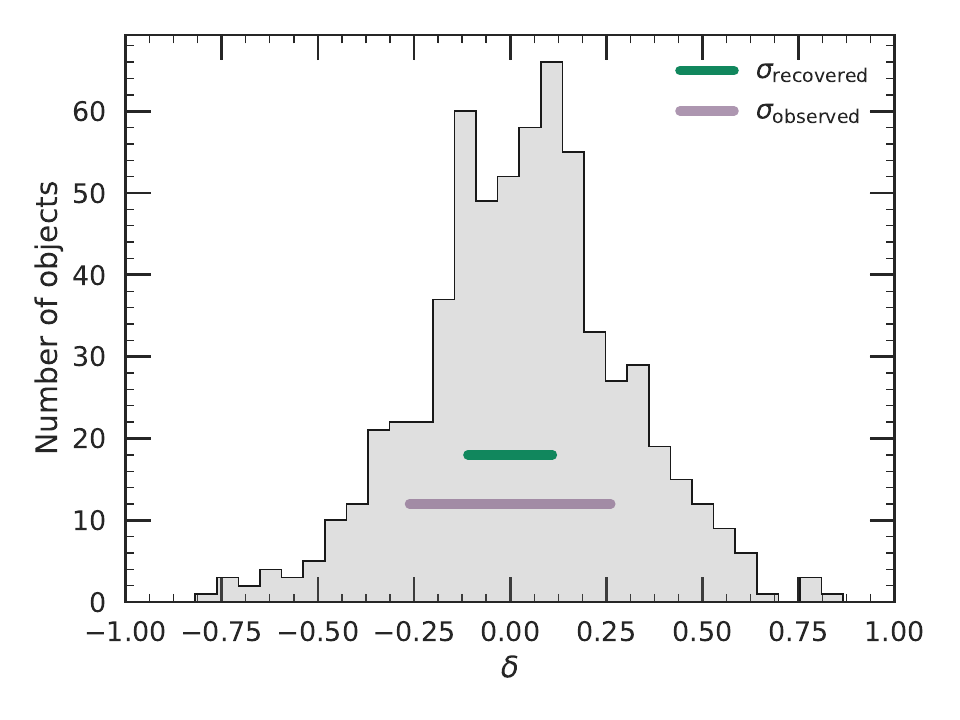}
  \caption{ Distribution of $\delta$ values for the $\sim 650$ objects measured in the COSMOS and GOODS-S fields, assuming a four age-bin SFH, a \citet{noll+09} attenuation law, and allowing stellar metallicity to be a free parameter. The standard deviation of the distribution is shown, along with the measurement error derived from our simulations.
  }
  \label{fig:delta-distribution}
\end{figure}

\begin{figure*}
  \captionsetup[subfigure]{labelformat=parens}
  \centering
  \subfloat[][]{\label{fig:delta-KDE}%
    \includegraphics[width=0.5\linewidth]{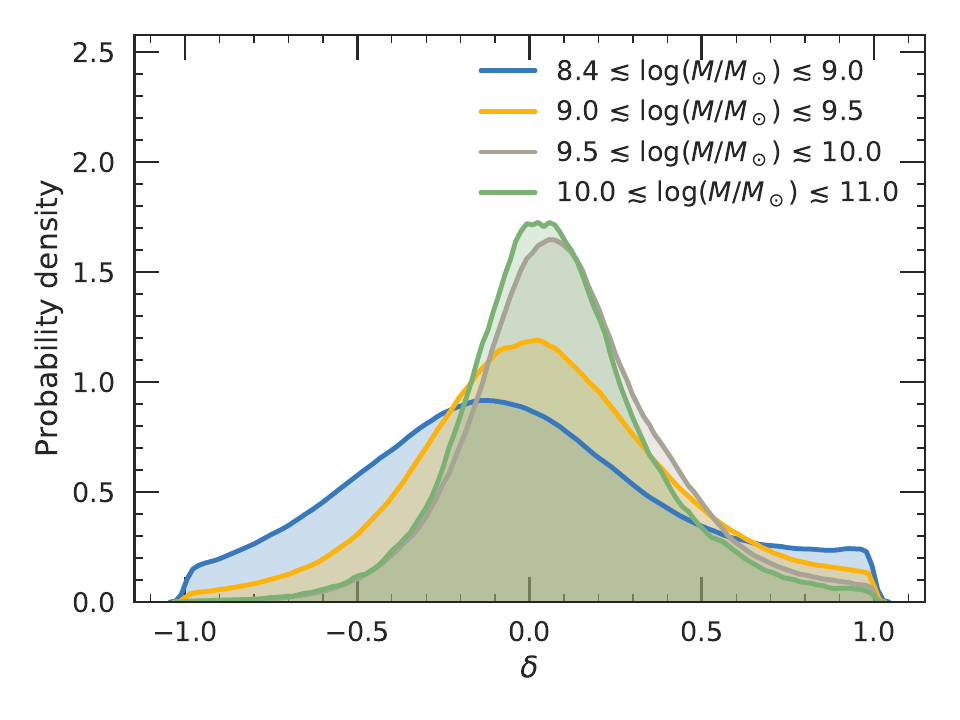}%
  }\hfill
  \subfloat[][]{\label{fig:Eb-KDE}%
    \includegraphics[width=0.5\linewidth]{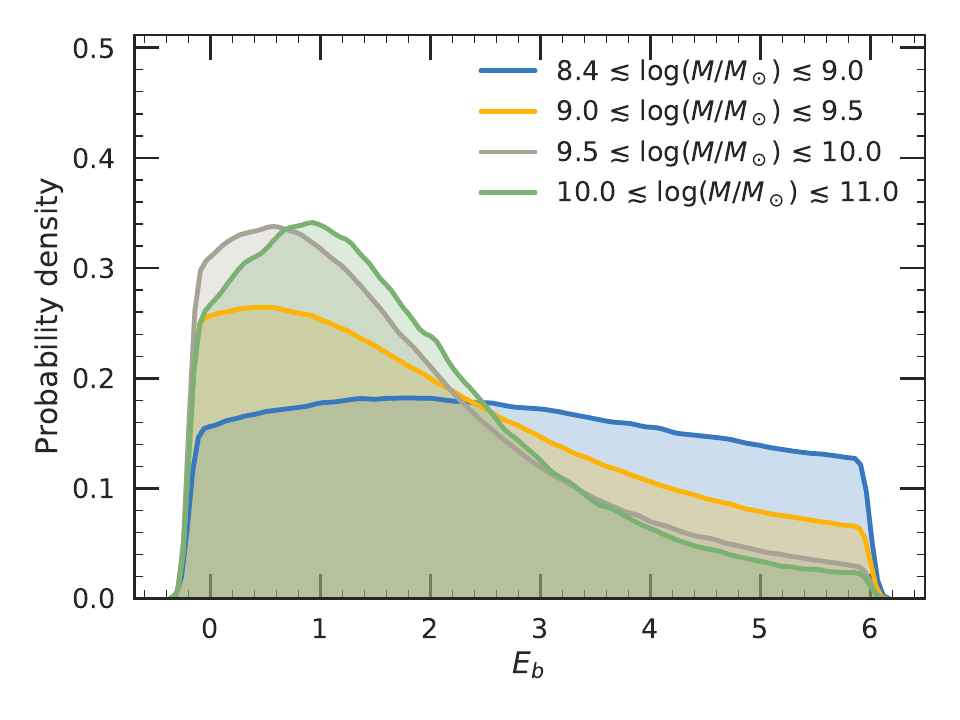}%
  }
  \caption{KDE curves for the full posterior distributions of $\delta$ (left) 
  and $E_b$ (right) across the four stellar mass-bins.  We find no significant correlation between the slope of the attenuation curve and stellar mass; the narrowing of the distributions towards higher mass are likely due to higher signal-to-noise in the photometry and an increase in the overall dust content, both of which allow for a more precise estimate of the dust law parameters.  The probability distributions for $E_b$ are much broader and exhibit a similar increasing precision towards higher masses. The low-mass bin just reflects the prior on the parameter while the higher-mass bins show increasing sensitivity to the variable.
  }
  \label{fig:dust-KDEs}
\end{figure*}

In the AEGIS, UDS, and GOODS-N fields, the tests discussed in Appendix~\ref{appendix:b} demonstrate that \mcsed\ can recover $\delta$ to a precision of $\sigma \lesssim 0.15$. In COSMOS and GOODS-S, the results are even better:  in these fields, the extensive photometric coverage in the rest-frame enables $\delta$ to be measured to $\sigma \lesssim 0.11$.  Figure~\ref{fig:delta-distribution} places this smaller dispersion in context by comparing it to the observed distribution of $\delta$ values.  Clearly, the spread in $\delta$ for the galaxies in the COSMOS and GOODS-S fields is broader than would be expected solely from measurement error, implying a physical variation in the intrinsic slope of the attenuation curve. 
Figure~\ref{fig:delta-KDE} shows the full probability distributions for $\delta$ across the four stellar mass-bins. While several studies have found that the slope of the attenuation curve changes with stellar mass \citep{zeimann+15b, buat+12, kriek+13}, the probability distributions for our fits do not show a significant correlation between the two properties.

Our \mcsed\ analysis also suggests that the 2175~\AA\ extinction bump may be present in the 
SEDs of $z \sim 2$ star-forming emission-line galaxies.  Using the dust models of \citet{silva+98}, \citet{granato+00} demonstrated that in local systems dominated by young stars, no 2175~\AA\ bump is visible. Conversely, in more ``normal'' star-forming galaxies, where a significant fraction of UV photons are emitted by stars outside their birth clouds, an extinction bump is present.  Hence the \citet{calzetti+94} observations of local starburst galaxies provide no evidence of a 2175~\AA\ extinction feature, while observations of less extreme systems \citep[e.g.,][]{burgarella+05, conroy+10, hoversten+11, battisti+17} exhibit the feature.

The 2175~\AA\ bump has been reliably measured in the stacked spectra of high-$z$ star forming galaxies \citep[e.g.,][]{kriek+13, zeimann+15b}. However, in contrast to $\delta$, our SED analysis is not particularly sensitive to $E_b$ (see Appendix~\ref{appendix:b}), and for individual galaxies, the bump is virtually undetectable. Figure~\ref{fig:Eb-KDE} shows the probability distributions of $E_b$ for the four stellar mass-bins and demonstrates the difficulty in constraining the strength of the dust bump. 

The main reason for the poor constraints on the 2175~\AA\ bump is the nature of the photometric data being analyzed. Most of the filters used to survey the CANDELS fields have bandpasses that are significantly broader than the 350~\AA\ width of the extinction feature.  Unless high signal-to-noise observations are taken through an intermediate bandpass filter centered near 2175~\AA\ rest frame and two bracketing filters, an accurate measurement of $E_b$ is nearly impossible.

\subsection{Spectrophotometric SED fitting}
\mcsed\ allows us to explore the degree to which our SED fits and parameter estimates are improved by including emission-line fluxes as input data. It is well-known that emission lines can provide information about a galaxy's current star-formation rate, metallicity, and warm ISM (density, pressure, and ionization parameter) that photometric measurements are unable to access \citep[e.g.,][]{ xiao+18, kewley+19}.  Moreover, Euclid \citep{NISP} and WFIRST \citep{WFIRST} will soon make the availability of such data commonplace, providing a new resource for understanding the properties of star-forming galaxies at high-$z$.  The question is, how much do the bright emission-lines of \OIII\ $\lambda\lambda 4959,5007$, \OII\ $\lambda 3727$, and H$\beta$ improve our ability to recover information about the physical properties of these galaxies.   In what follows, nebular emission (both continuum and line emission) is included in all our SED fits; the difference is only whether the observed emission line fluxes \citep[courtesy of \threedhst;][]{3DHST} are employed to constrain the models.

\mcsed\ handles emission lines in an analogous way to photometry.  Input line fluxes are compared to the predictions of a grid of \cloudy\ models, and the resultant $\chi^2$ term contributes to the goodness-of-fit of the SED\null.  Because some emission lines are more difficult to model than others, \mcsed\ allows users to weight each line's contribution. In our case, we weight the \Hb\ line measurement equal to that of a photometric measurement and give half-weight to the strengths of the  collisionally-excited of \OII\ and \OIII, since such features depend as much on the physical conditions of the ISM as they do on ionization flux of bright stars \citep[\eg][]{kewley+02, shapley+15}. Although the precise weights given to these emission lines has minimal impact on the overall results, this particular weighting scheme typically produces better $\chi^2$ values than other formulations.

\begin{figure*}[ht!]
  \captionsetup[subfigure]{labelformat=empty}
  \centering
  \subfloat[][]{\label{}%
    \includegraphics[width=0.5\linewidth]{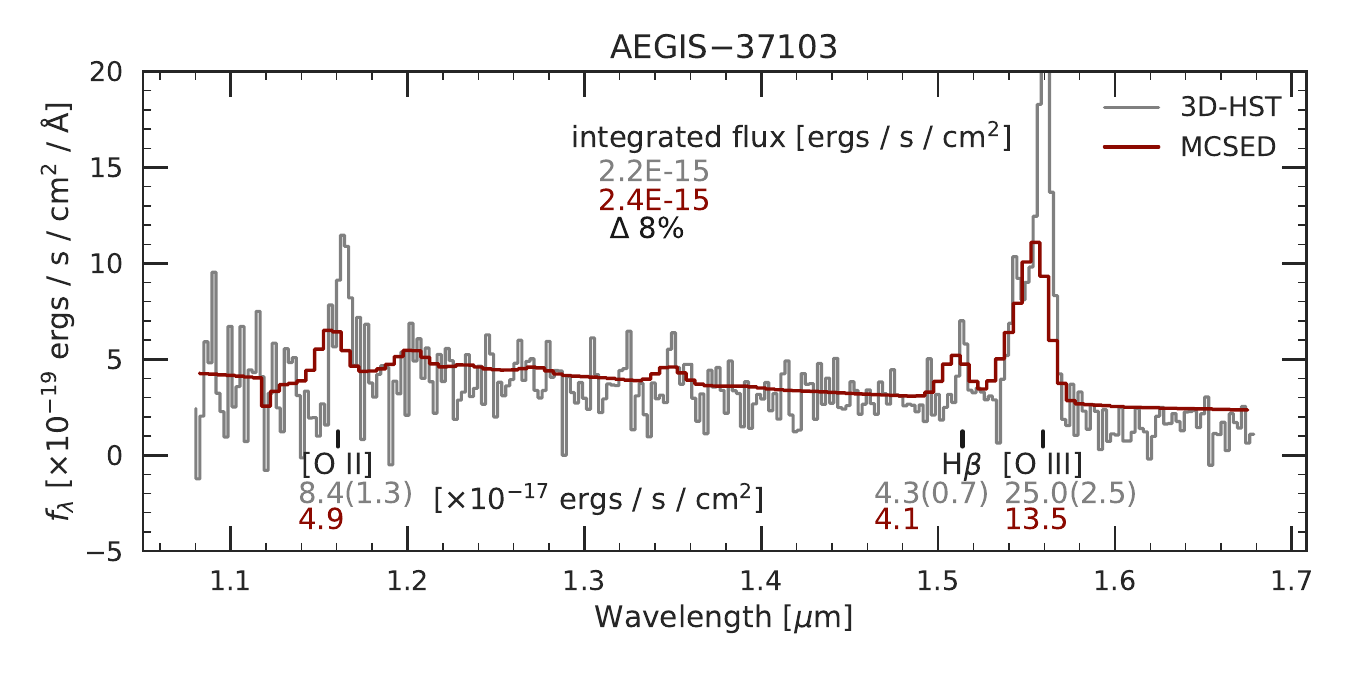}%
  }
  \subfloat[][]{\label{}%
    \includegraphics[width=0.5\linewidth]{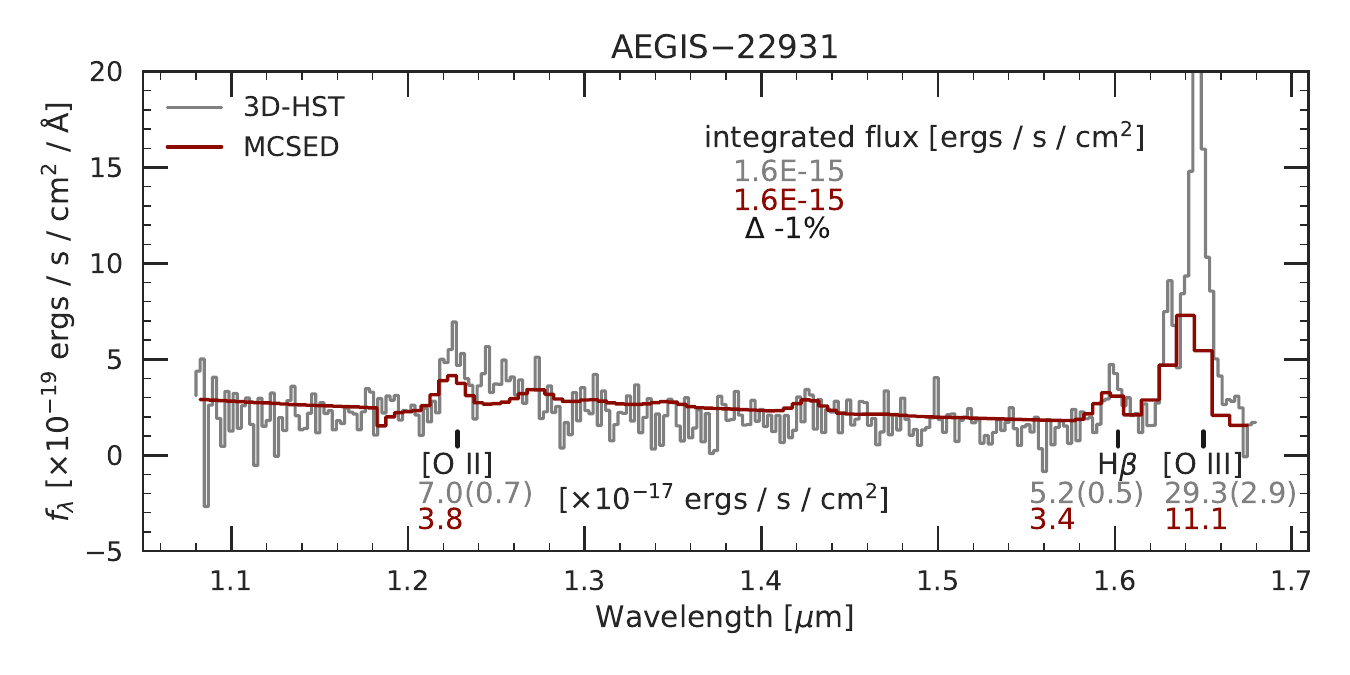}%
  }
  \\[-8ex]
  \subfloat[][]{\label{}%
    \includegraphics[width=0.5\linewidth]{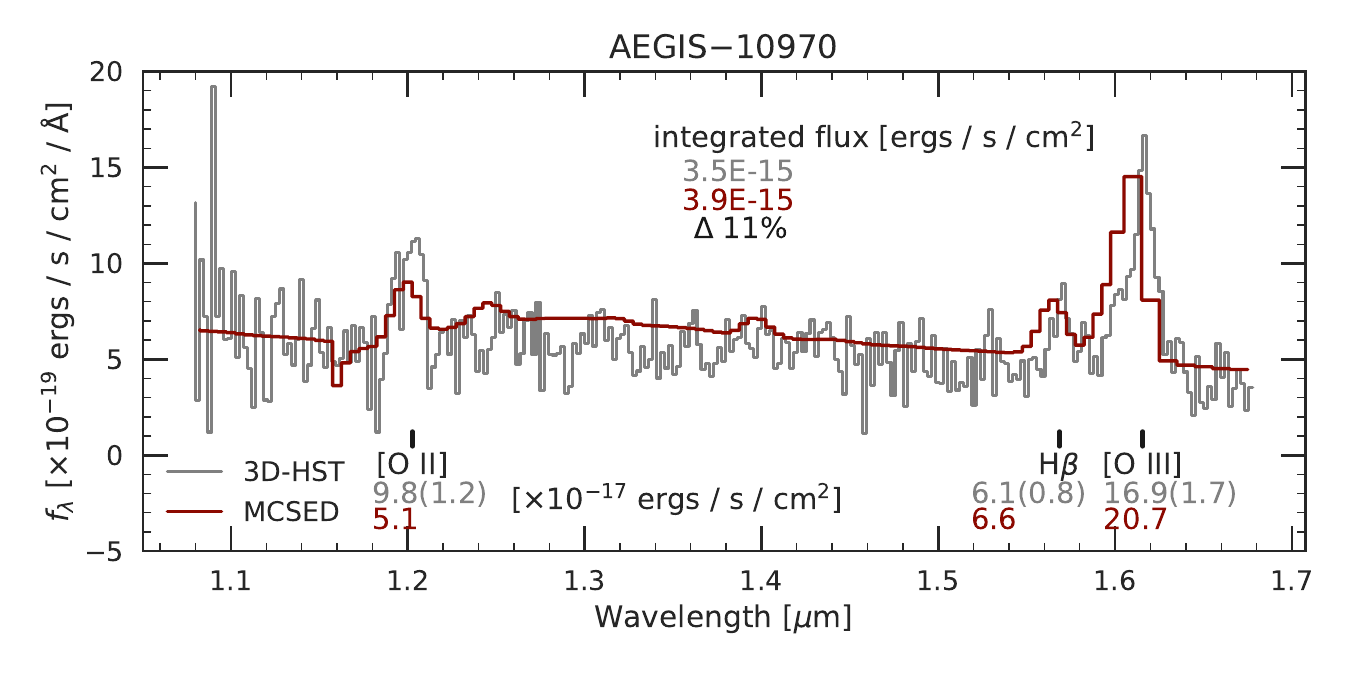}%
  }
  \subfloat[][]{\label{}%
    \includegraphics[width=0.5\linewidth]{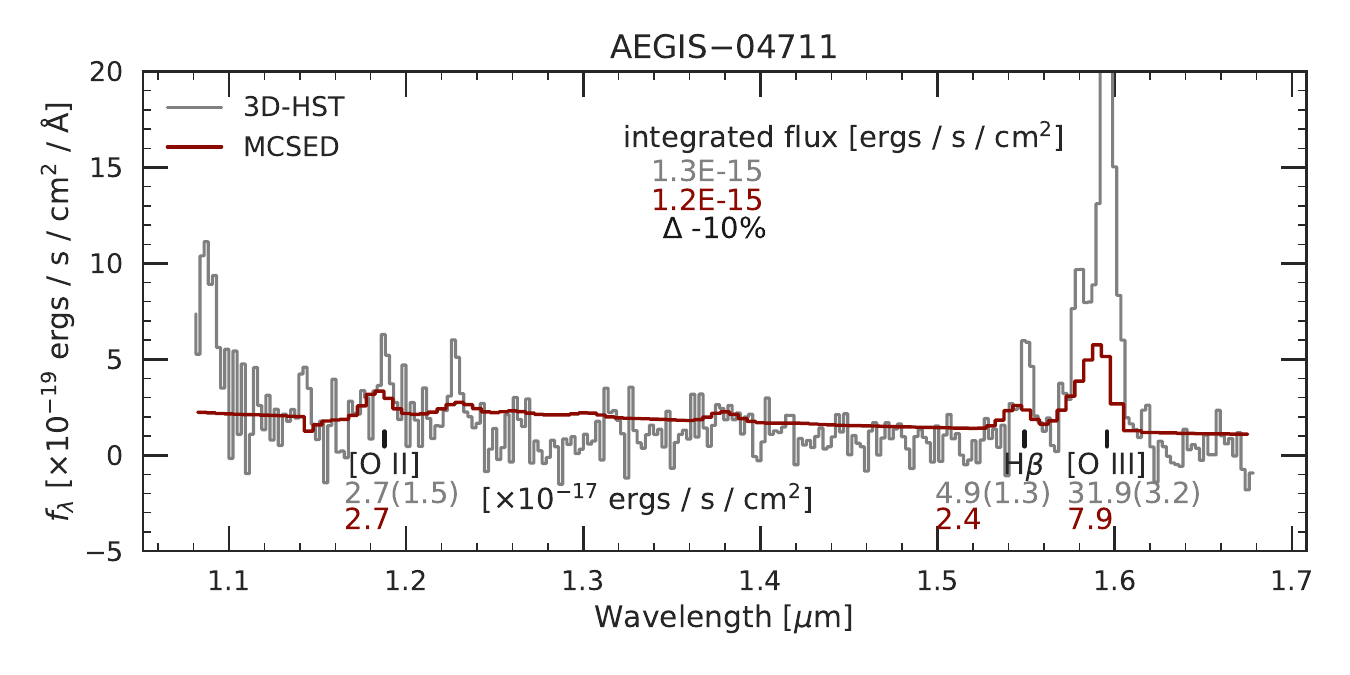}%
  }
  \\[-8ex]
  \subfloat[][]{\label{}%
    \includegraphics[width=0.5\linewidth]{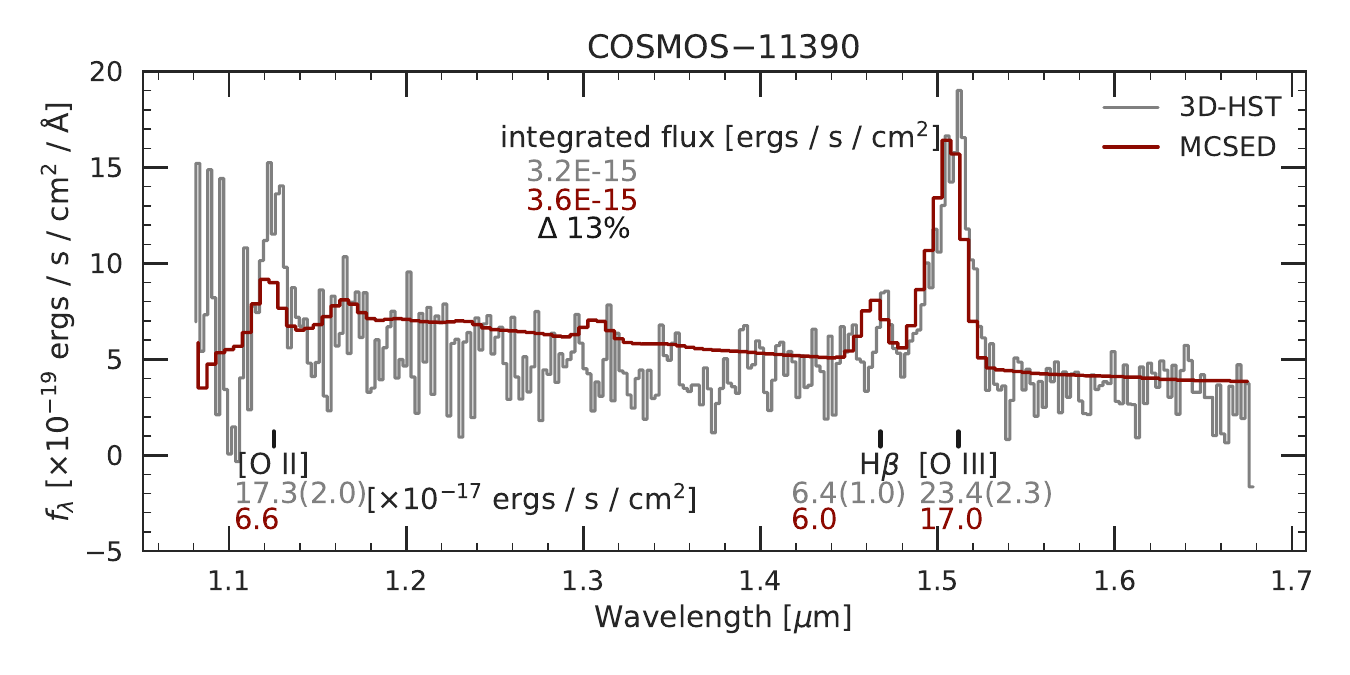}%
  }
    \includegraphics[width=0.5\linewidth]{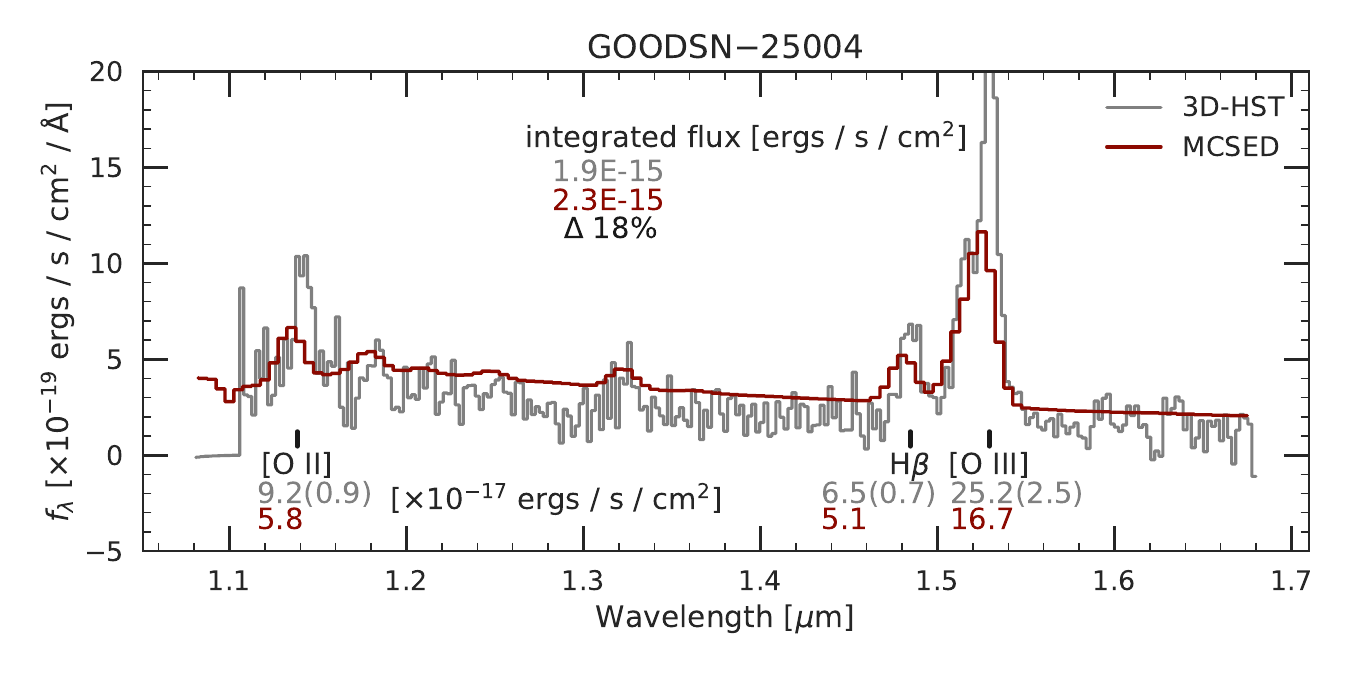}
  \\[-8ex]
  \subfloat[][]{\label{}%
    \includegraphics[width=0.5\linewidth]{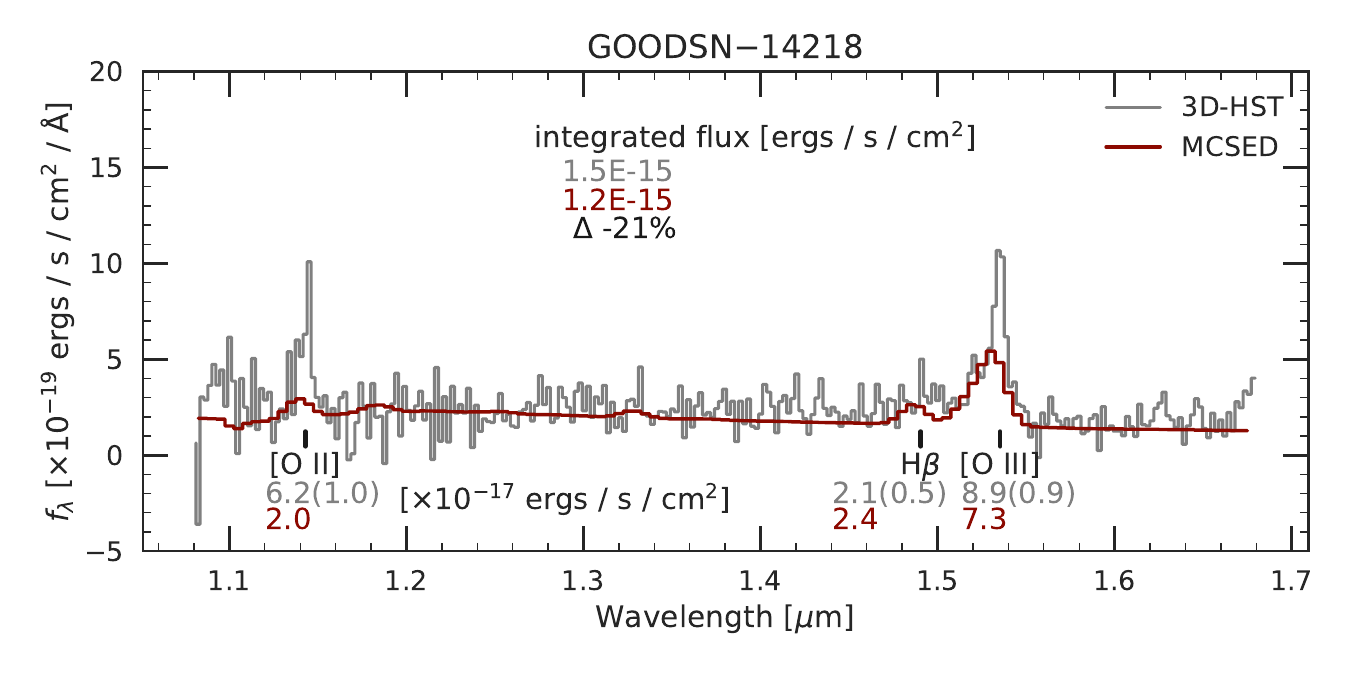}%
  }
  \subfloat[][]{\label{}%
    \includegraphics[width=0.5\linewidth]{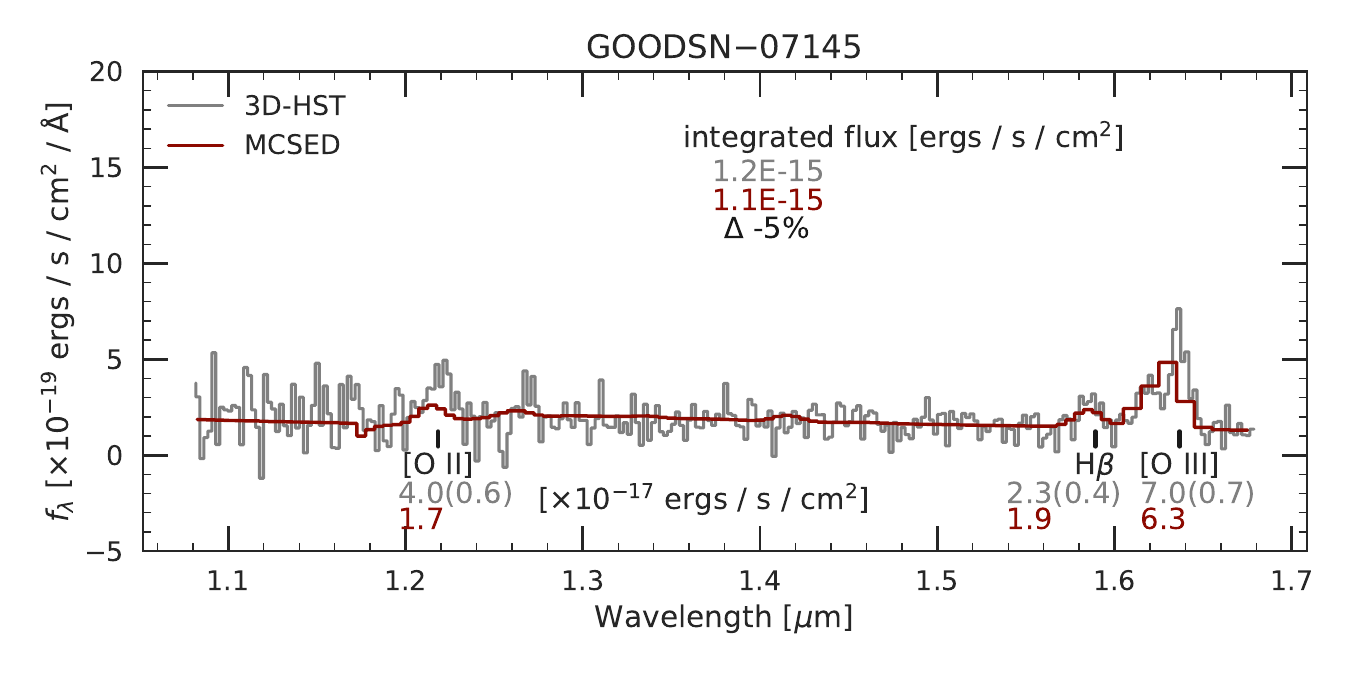}%
  }
  \caption{Comparison of the modeled SEDs (red) to the observed spectra (grey) for 8 emission-line galaxies in the \threedhst\ survey.  
  The \Hb\ line fluxes are generally well-modeled, while the collisionally-excited oxygen lines are typically underpredicted. (The \OIII\ fluxes printed on the panels refer to the $\lambda 5007$ line alone.) The integrated fluxes withing the \threedhst\ bandpass are generally consistent to within $\sim20\%$.
  }
  \label{fig:cf-3dhst-mcsed-spectra}
\end{figure*}

\begin{figure*}[ht!]
  \captionsetup[subfigure]{labelformat=parens}
  \centering
  \noindent\includegraphics[width=\linewidth]{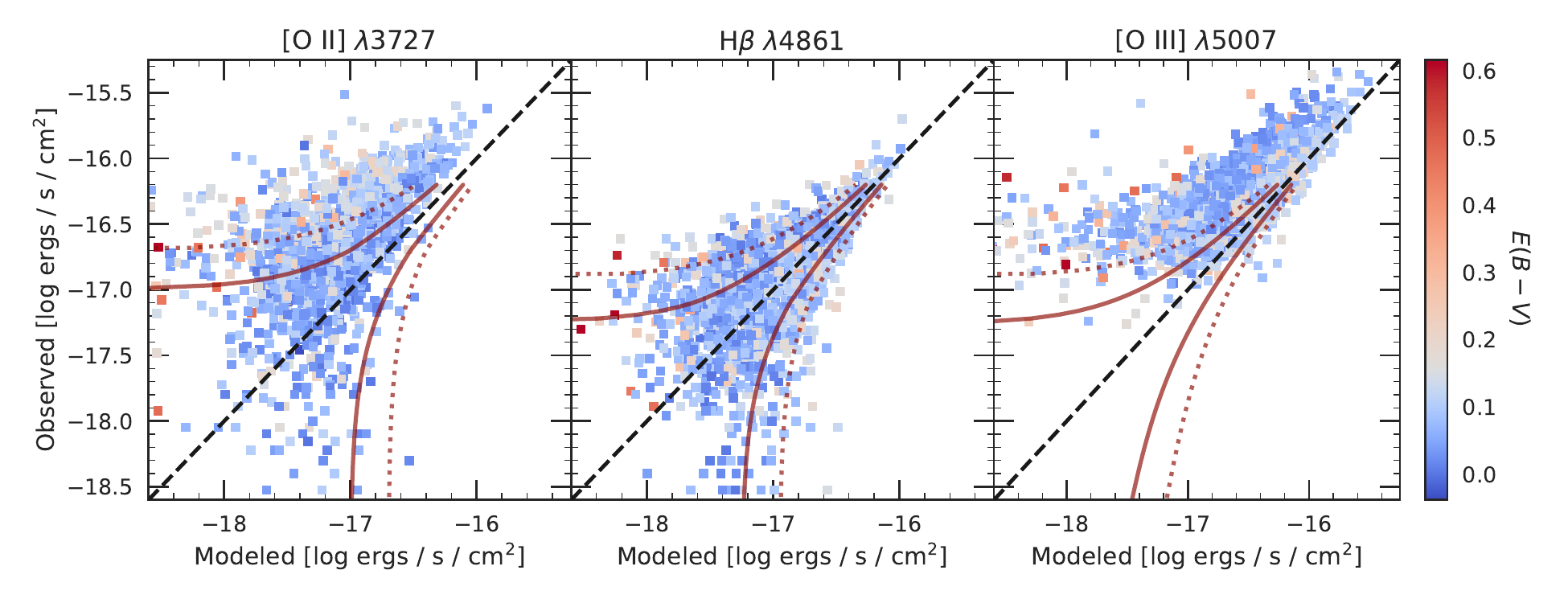}
  \caption{ Comparison of the modeled and observed line fluxes for \OII\ $\lambda 3727$, \Hb, and \OIII\ $\lambda 5007$. \Hb\ shows excellent agreement, while the strengths of the collisionally-excited oxygen lines are systematically underpredicted. The solid (dotted) red line shows the scatter about the 1:1 relation within the 68\% (95\%) confidence interval that is expected solely from uncertainties in the line flux measurements. Objects with larger internal reddening are more susceptible to offsets in the modeled vs.\ measured line fluxes, but that is not the only (or dominant) effect contributing to the discrepancy.}
  \label{fig:cf-3dhst-mcsed-lineflux}
\end{figure*}

Before investigating the impact that emission lines have on our fits, we first compare our model SEDs and line strengths to observations from the \threedhst\ survey \citep{3DHST}. Figure~\ref{fig:cf-3dhst-mcsed-spectra} shows several examples of this comparison and Figure~\ref{fig:cf-3dhst-mcsed-lineflux} plots the modeled fluxes against the observed values.

\begin{figure}[h!]
  \captionsetup[subfigure]{labelformat=parens}
  \centering
  \noindent\includegraphics[width=\linewidth]{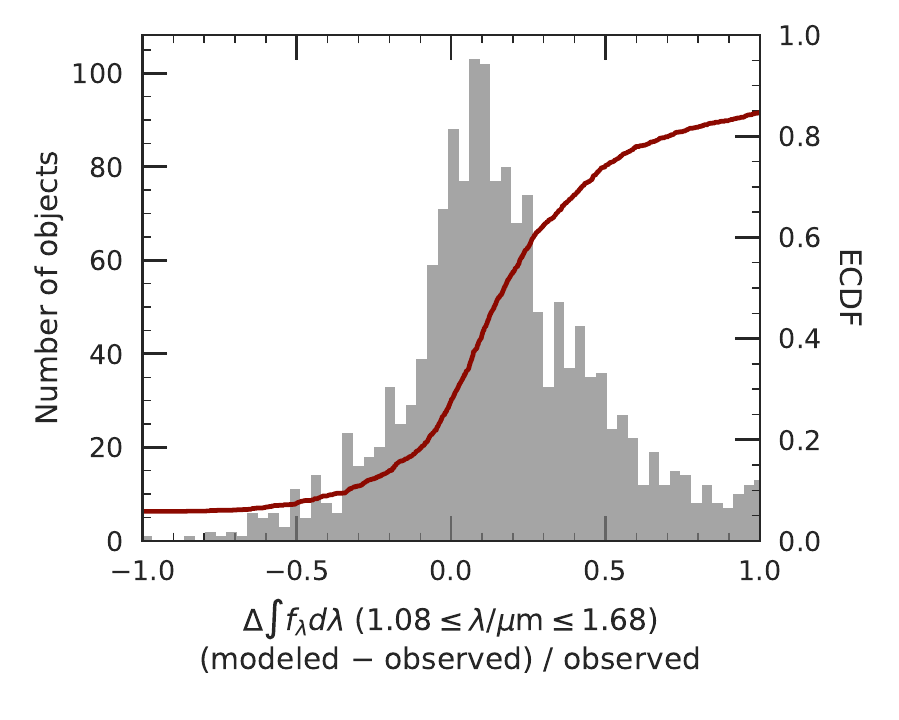}
  \caption{ Comparison of the modeled and observed fluxes integrated over the \threedhst\ spectral range ($1.08~\mu{\rm m} < \lambda < 1.68~\mu{\rm m}$). The fractional change in the integrated flux is calculated as the difference between the modeled (\mcsed ) and observed (\threedhst ) flux, normalized by the observed flux. The systematic underprediction of the \OII\ and \OIII\ emission-line fluxes is compensated for by a mild overestimate of the surrounding continuum. This brings the integrated fluxes predicted by the models into good agreement with the photometry.}
  \label{fig:cf-3dhst-mcsed-integrated-flux}
\end{figure}

Both figures demonstrate that the modeled strength of \Hb\ agrees very well with the flux estimates from \threedhst. Given the relative insensitivity of the hydrogen recombination lines to the physical conditions of the ISM \citep[\eg][]{Osterbrock}, this agreement is not surprising. In contrast, the collisionally-excited \OII\ and \OIII\ lines are consistently underpredicted by our models, and to compensate for this underprediction, the surrounding stellar continuum is slightly overpredicted.  As a result, the modeled and observed fluxes integrated over the \threedhst\ spectral range are in good agreement (typically within $10-20\%$; see Figure~\ref{fig:cf-3dhst-mcsed-integrated-flux}) and do not exhibit any obvious systematic behavior.

\begin{figure*}[ht!]
  \captionsetup[subfigure]{labelformat=parens}
  \centering
  \subfloat[][]{\label{fig:ZlogU-OIIIHb}%
    \includegraphics[width=0.5\linewidth]{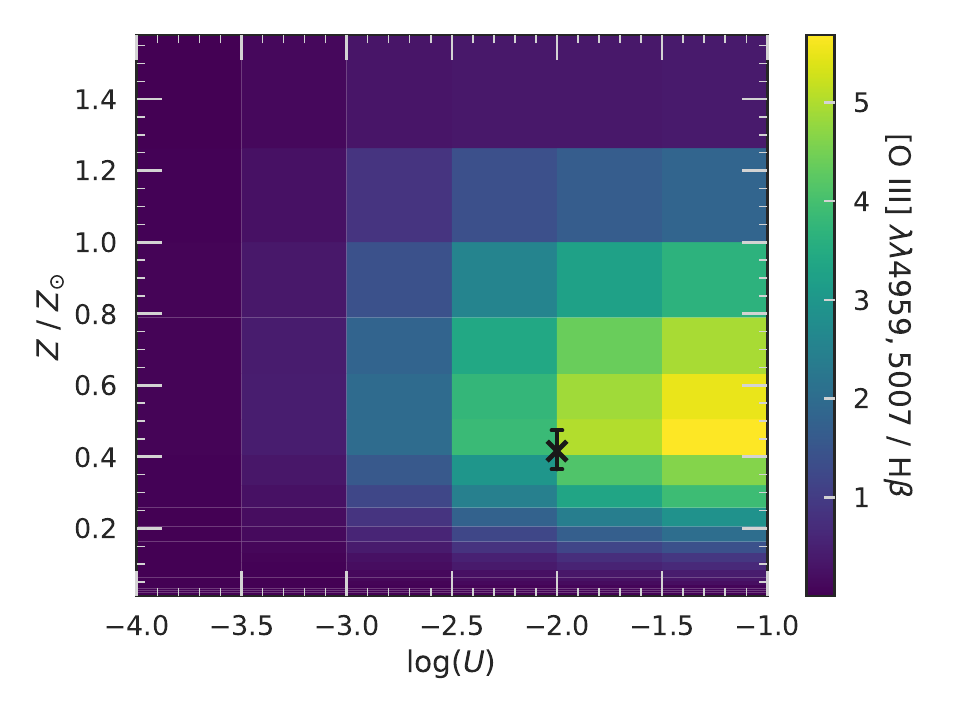}%
  }\hfill
  \subfloat[][]{\label{fig:OIIIHb-observed}%
    \includegraphics[width=0.5\linewidth]{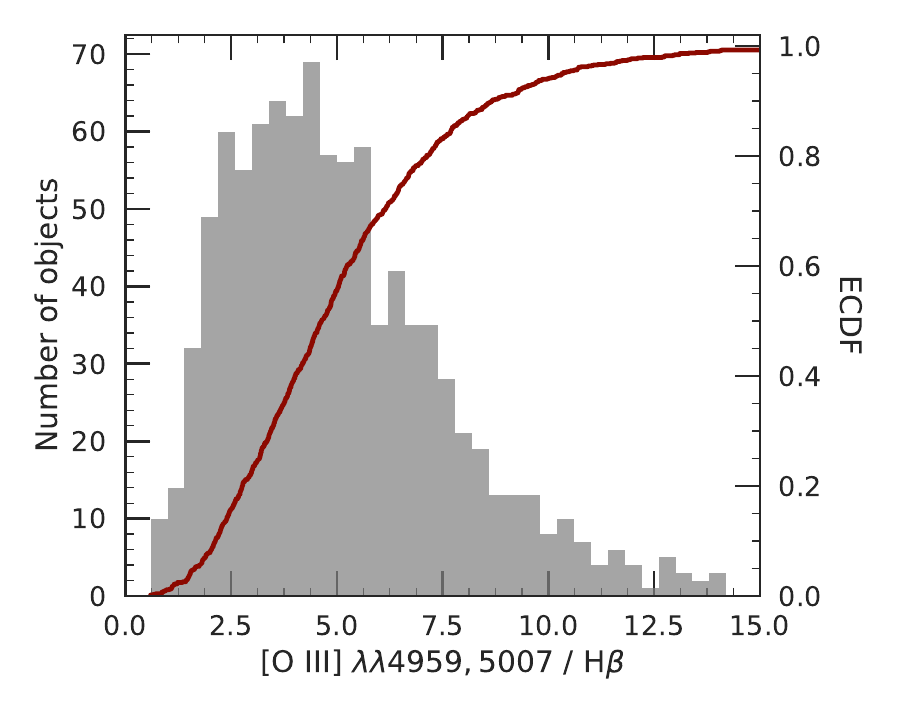}%
  }
  \caption{{\it Left:} The variation of the \OIII\ $\lambda\lambda 4959,5007$/\Hb\ flux ratio (from the nebular emission models) as a function of ionization parameter ($x$-axis) and metallicity ($y$-axis).  In our analysis, we fixed the ionization parameter ($\log U=-2$) and treated metallicity as a free parameter.  The error bar gives the best-fit metallicity ($\sim 42\%$) solar and 68\% uncertainty across our entire galaxy sample.  
 {\it Right:} The distribution of \OIII\ $\lambda\lambda4959,5007$/\Hb\ flux ratios measured for our sample of \threedhst\ emission-line galaxies. The peak of the distribution is near $\sim 4$ but a significant fraction of galaxies ($\gtrsim 30\%$) have line ratios that exceed those allowed by the nebular emission models. Only those objects with well-measured emission-lines are shown; the precise criteria used to exclude objects with uncertain line ratios have no effect on the distribution.  Known AGN have been excluded from the sample.  Although some elevated line ratios may be due to low-luminosity AGN, a stacking analysis of deep GOODS-S X-ray data suggest the total fraction of AGN to be less than $\sim 5\%$ \citep{bowman+19}.
  }
  \label{fig:OIII-Hb-ratios}
\end{figure*}

The primary reason that our SED fits consistently underpredict \OIII\ $\lambda 5007$ is that none of the \cloudy-based models generated by \citet{byler+17} produce the elevated line ratios that are observed by \threedhst.  This behavior is illustrated in  Figure~\ref{fig:ZlogU-OIIIHb}, which displays the ratio of \OIII\ $\lambda\lambda 4959,5007$ to \Hb\ as a function of metallicity and ionization parameter.  None of the models have \OIII/H$\beta$ ratios greater than 6.  Yet as is evident from Figure~\ref{fig:OIIIHb-observed}, the typical \OIII/H$\beta$ ratio seen in our 
sample of $z \sim 2$ emission-line galaxies is $\sim 4$, and the distribution exhibits an extended tail towards more elevated ratios.  This is not a new result:  high \OIII/\Hb\ ratios have been widely observed in the $z\gtrsim1$ universe \citep[e.g.,][]{maseda+14, steidel+14, dickey+16}, but the reason behind the phenomenon has yet to be fully understood \citep[see][and references therein]{kewley+19}.

As noted by \citet{byler+17}, other combinations of ionizing spectra and photoionization codes, such as those presented in \citet{dopita+13} using STARBURST99 \citep{STARBURST99} and \mbox{MAPPINGS III} \citep{MAPPINGSIII} can produce \OIII/\Hb\ ratios that more closely resemble our sample. Similarly, population synthesis models that include binary evolution \citep[i.e., {\tt BPASS};][]{BPASS} have also proven successful at reproducing elevated \OIII/\Hb\ line ratios \citep{stanway+14}.  Nevertheless, the \citet{byler+17} grid of \cloudy\ models do have the advantage of self-consistency, as they employ the same FSPS SEDs that we use to model the galaxies' broadband colors. Thus, in what follows, we use the \citet{byler+17} grid, accepting that their collisionally-excited line fluxes are systematically smaller than the observed values.

(We note that AGN are known to play a role in creating \OIII\ to H$\beta$ ratios \citep[e.g.,][]{BPT, shapley+15}, and, as currently configured, \mcsed\ does not model emission this emission. However, \citet{bowman+19} already removed the vast majority of AGN from the sample and, based on the exceedingly deep X-ray data of the Chandra Deep Field South, estimate the fraction of low-luminosity AGN to be no more than $\sim 5\%$.  Hence this source of emission should not be affecting our line ratios.)

Our rest-frame optical emission line measurements primarily constrain three physical properties.  The first, the ISM's ionization parameter, is implicitly taken into account by our choice of assumptions.  As noted above, the nebular models cannot reproduce the elevated line ratios that are ubiquitous in the high-redshift universe.  Nonetheless, as Figure~\ref{fig:ZlogU-OIIIHb} illustrates, the high value of ionization parameter that we adopt here ($\log U = -2$) does a reasonable job of fitting the high-excitation line ratios that are exhibited by our sample.

The second property constrained by our emission-line measurements is star formation.  All three of our fitted emission lines provide some indication of the flux of ionizing photons currently being produced within a galaxy, and this quantity is directly tied to the very recent ($t \lesssim 10$~Myr) rate of star formation \citep[e.g.,][]{kennicutt98, kennicutt+12}.  However, each line has its drawbacks. \Hb\ is most directly tied to star formation, since for Case~B recombination, Balmer-line fluxes directly measure the photoionization rate.  However, in the majority of our $z \sim 2$ galaxies, \Hb\ is only marginally detected ($\sim 60$\% have a signal-to-noise ratio less than 2) and, at a rest-frame wavelength of 4861~\AA, the effects of attenuation cannot be ignored.  Alternatively, one can translate the collisionally-excited oxygen lines into a SFR estimate.  However, in both the local \citep{moustakas+06} and distant \citep{teplitz+00} universe, the scatter between \OIII\ and Balmer-line based SFRs is more than a factor of 2, and, although \OII\ $\lambda 3727$ is often used as a local SFR indicator \citep{kewley+04, moustakas+06, kennicutt+12}, the line is generally weak at high redshift and is sensitive to the affects of both metallicity and dust.  Nevertheless, when taken together, these lines do provide a measure of recent SFR that complements that based on light from the stellar continuum.

\begin{figure*}[hp!]
  \captionsetup[subfigure]{labelformat=parens}
  \centering
  \subfloat[][]{\label{fig:emline-kde-sfr}%
    \includegraphics[width=0.33\linewidth]{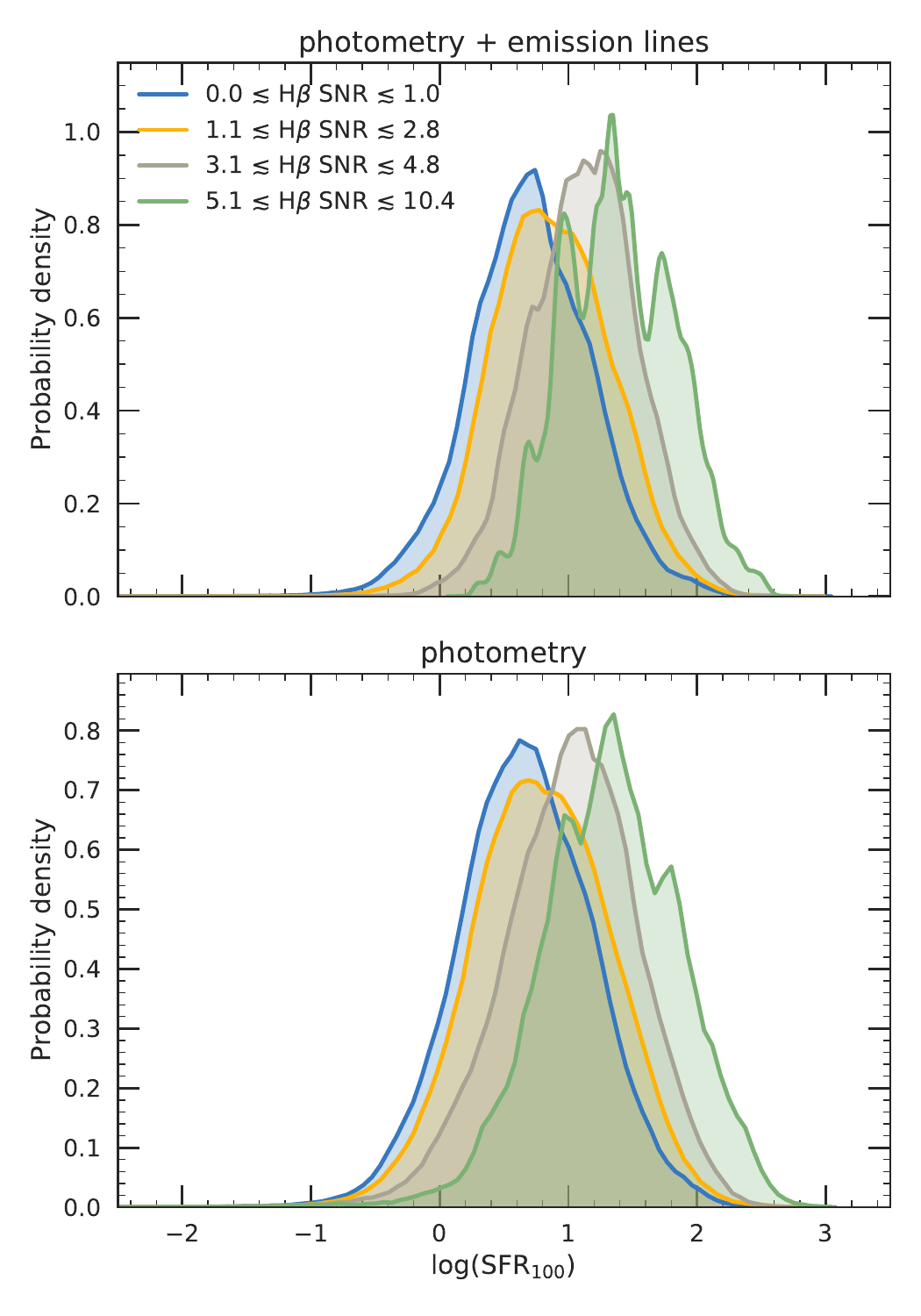}%
  }\hfill
  \subfloat[][]{\label{fig:emline-kde-EBV}%
    \includegraphics[width=0.33\linewidth]{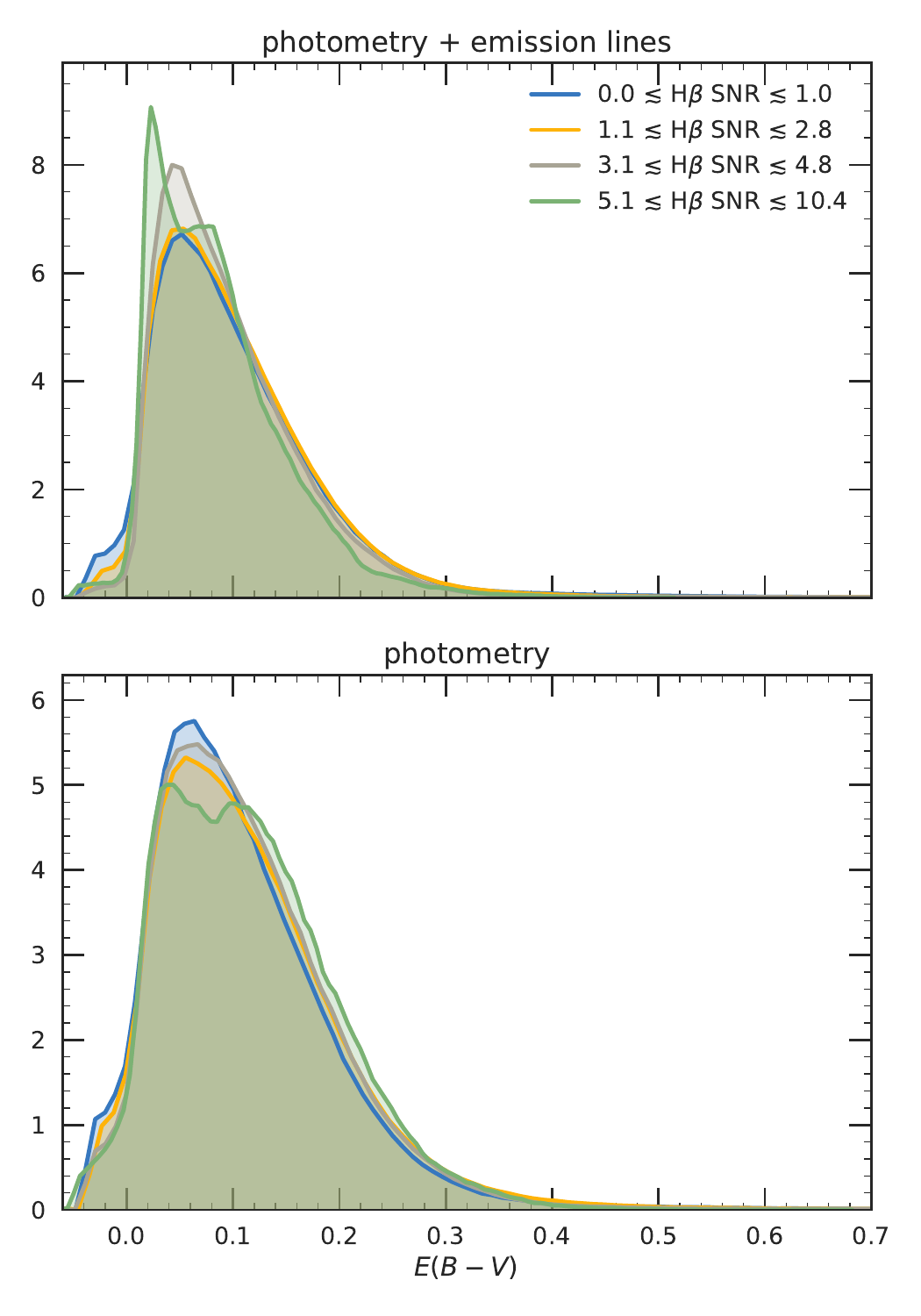}%
  }\hfill
  \subfloat[][]{\label{fig:emline-kde-logZ}%
    \includegraphics[width=0.33\linewidth]{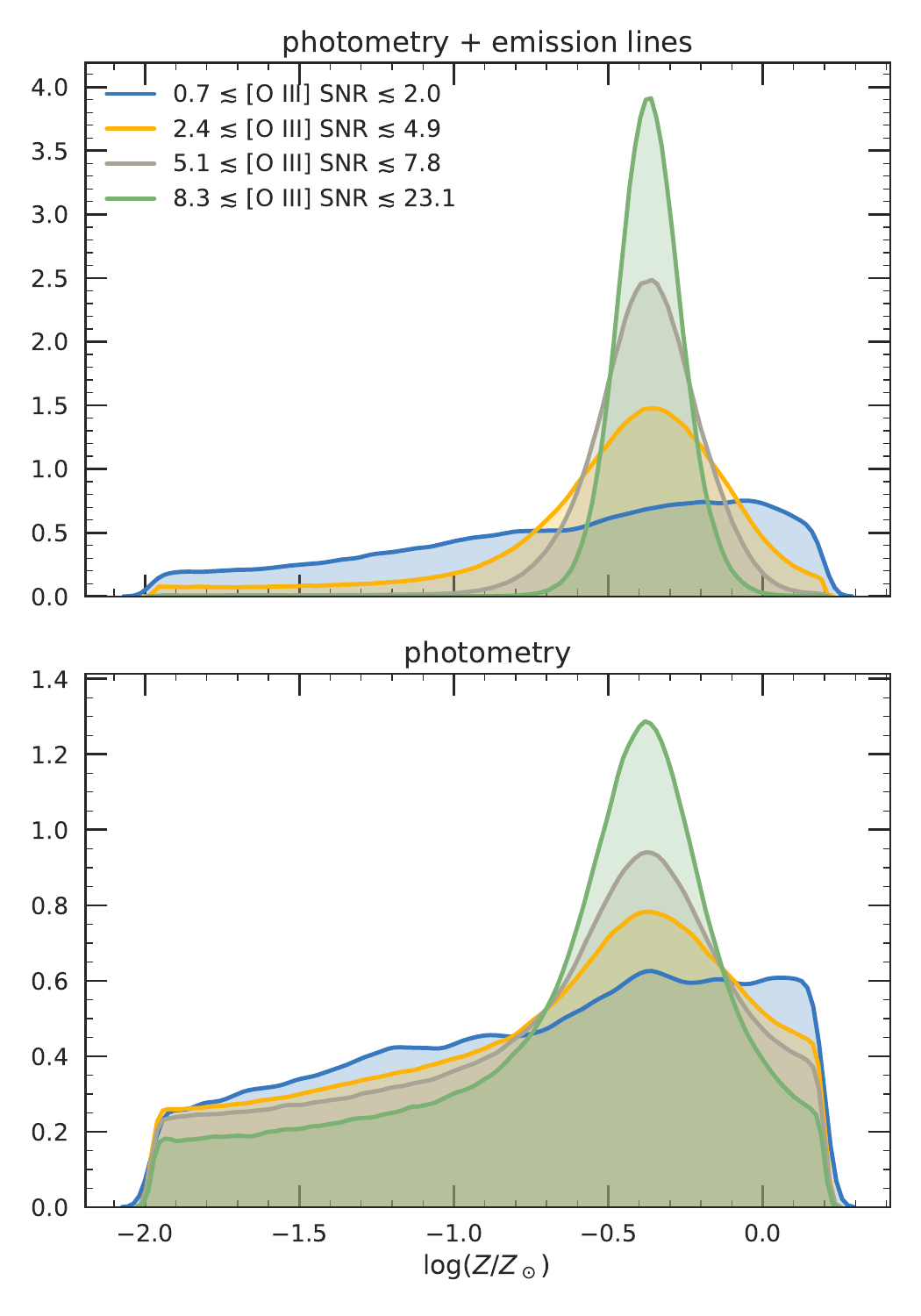}%
  }
  \caption{Comparison of posterior probability distributions of the SFR over the past 100~Myr,  $E(B-V)$, and $\log(Z/Z_\odot)$, with and without the inclusion of observed emission line fluxes in the $\chi^2$ calculation for the SED fits.  Emission lines produce only a marginal improvement in the measurement of the current star formation rate, as the wealth of rest-frame UV photometry already provides tight constraints on this parameter.  The improvement in $E(B-V)$ is similarly small, as it results mostly from the strong degeneracy between star formation and extinction (\ie a more precise estimate of recent star formation yields a more precise estimate of the total dust content). However, the inclusion of emission lines greatly improves our estimate of metallicity, as the line ratios provide insight on the nebular oxygen abundance (and, by direct extension, the present-day stellar metallicity). The probability distributions remain broad when the emission lines are only weakly measured, but become significantly narrower in the set of objects with high signal-to-noise measurements.
  }
  \label{fig:cf-bestfit-w-wo-emlines}
\end{figure*}

\begin{figure*}[hp!]
  \captionsetup[subfigure]{labelformat=parens}
  \centering
  \subfloat[][]{\label{fig:emline-kde-sfr-centered}%
    \includegraphics[width=0.33\linewidth]{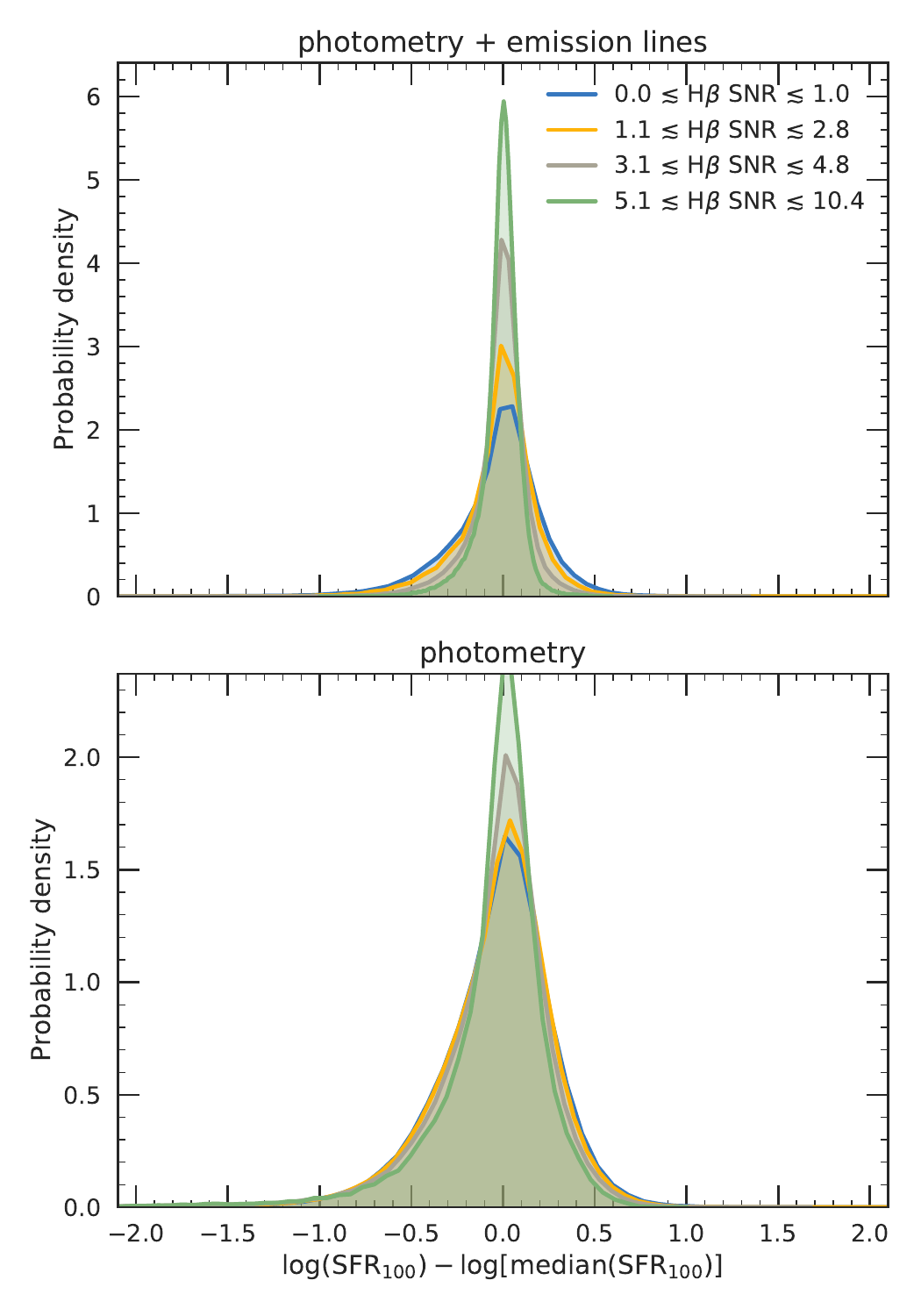}%
  }\hfill
  \subfloat[][]{\label{fig:emline-kde-EBV-centered}%
    \includegraphics[width=0.33\linewidth]{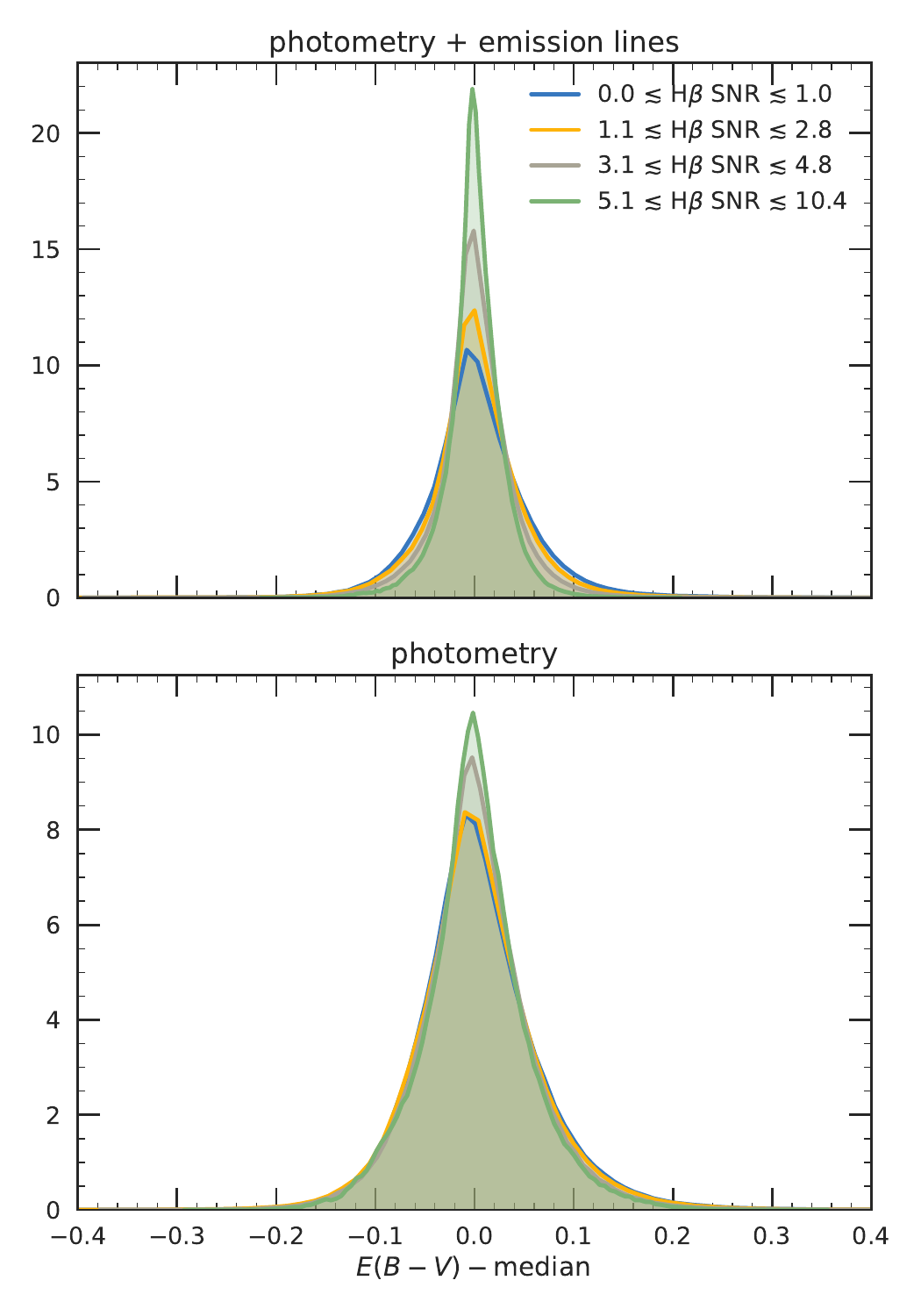}%
  }\hfill  
  \subfloat[][]{\label{fig:emline-kde-logZ-centered}%
    \includegraphics[width=0.33\linewidth]{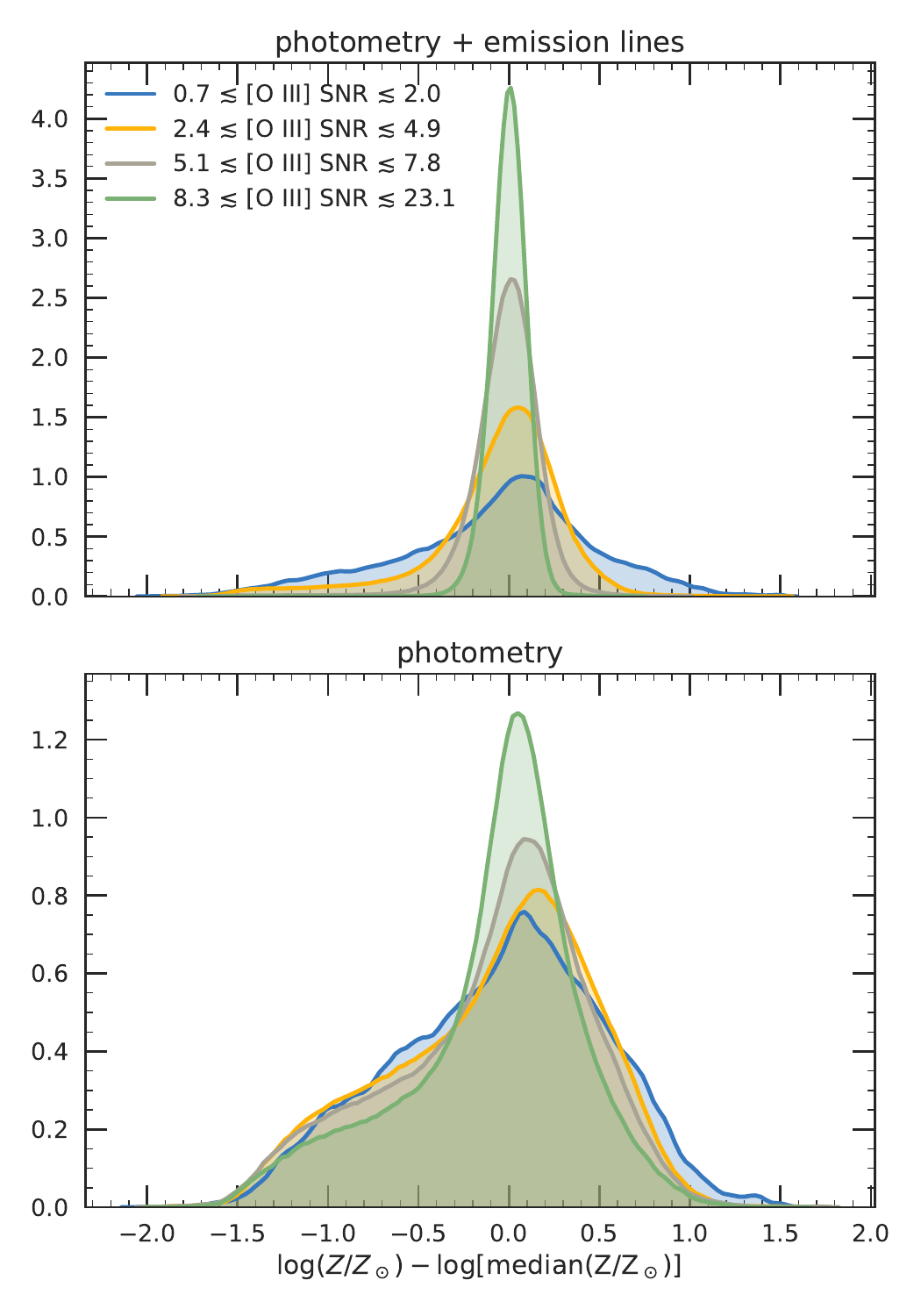}%
  }
  \caption{The same distributions as in Figure~\ref{fig:cf-bestfit-w-wo-emlines}, this time recentered about the peaks of the distributions.  Including observed emission line fluxes in the $\chi^2$ calculation (top row of panels) yields tighter constraints on recent star formation, dust content, and stellar metallicity.
  }
  \label{fig:cf-bestfit-w-wo-emlines-centered}
\end{figure*}

Figures~\ref{fig:emline-kde-sfr} and \ref{fig:emline-kde-sfr-centered} demonstrate this agreement by showing how the posterior probability distributions are narrower when well-measured emission lines are included as inputs to the fits.  By extension, the inclusion of emission lines also helps to constrain $E(B-V)$ (Figures~\ref{fig:emline-kde-EBV} and \ref{fig:emline-kde-EBV-centered}), as they help break the strong degeneracy between star formation and dust attenuation (see Figure~\ref{fig:triangle}). The largest improvement is exhibited in systems where the emission lines have the highest signal-to-noise ratios.

The third galaxy property improved by the inclusion of emission lines is metallicity. While measurements of absorption line indices can constrain the metal abundance of a stellar population \citep[e.g.,][]{worthey94, maraston+11}, none of these features are strong enough to be detected via grism spectroscopy or broadband photometry. In contrast, emission-line ratios can provide strong constraints on the metal abundance of a galaxy's ISM, and, by extension, its current generation of stars.  Figures~\ref{fig:emline-kde-logZ} and \ref{fig:emline-kde-logZ-centered} demonstrate this effect.  When emission line fluxes are not included as inputs to the fits, the metallicity posterior probability distributions are extremely broad. In contrast, when emission line are included (and well-measured), these same distributions are strikingly narrow.  Interestingly, the inclusion of emission lines in the $\chi^2$ fits does not significantly affect where the peak of the distribution lies, suggesting that the use of broadband and intermediate-band photometry alone can do a reasonable job of estimating mean metallicity.  This behavior may be unique to galaxy samples with strong emission lines (e.g., in the case of our $z \sim 2$ sample, line fluxes can contribute up to $\sim 50\%$ of the total flux in a given filter) where the broadband photometry implicitly contains useful information about emission-line strengths.
Nonetheless, including emission-line fluxes in the $\chi^2$ calculation significantly improves the metallicity measurement.

\section{Fitting Galaxies in the Mid- and Far-IR}
\label{sec:farIR}

Mid- and far-IR dust emission in starbursting galaxies arises from the re-radiation of the UV stellar continuum, and thus can trace dust attenuation. Observationally speaking, however, this principle of energy balance may not always hold:  while the infrared emission from dust is roughly isotropic, an attenuation measurement is valid only for a specific line of sight \citep[e.g.,][]{hayward+15}.  This situation complicates any analysis that relies upon a galaxy's IRX, i.e., the infrared to ultraviolet flux ratio \citep[see][]{dacunha+08}.  Furthermore, since the only observables of a galaxy at $z \gtrsim 4$  may be the UV luminosity and spectral slope \citep[e.g.,][]{bouwens+09, finkelstein+15}, it is important to understand how well energy conservation and dust attenuation can predict the infrared luminosity of high-$z$ galaxies.

The \citet{bowman+19} galaxy sample is not ideal for answering this question.  At $z \sim 2$, these systems are generally too faint to have {\sl Herschel/}PACS and SPIRE far-IR measurements, and, even in the mid-IR, reliable photometric measurements are difficult, due to the blending of nearby sources induced by the large instrumental point-spread-functions at these long wavelengths.  Nevertheless, we can explore the capabilities of \mcsed\ by analyzing one galaxy whose photometry appears to be reliable.

We used the CANDELS images and data products of \citet{barro+19} to select a galaxy whose mid- and far-IR photometry appears to be relatively unaffected by blending issues and image confusion. We then ran \mcsed\ twice, only changing the treatment of dust mass.  For our first run, the dust mass was treated as a free parameter that set the normalization of the dust emission; for the second calculation, we assumed that all of the energy attenuated by dust was re-radiated in the IR (i.e., using the energy balance argument).  

\begin{figure}[h!]
  \centering
  \noindent\includegraphics[width=\linewidth]{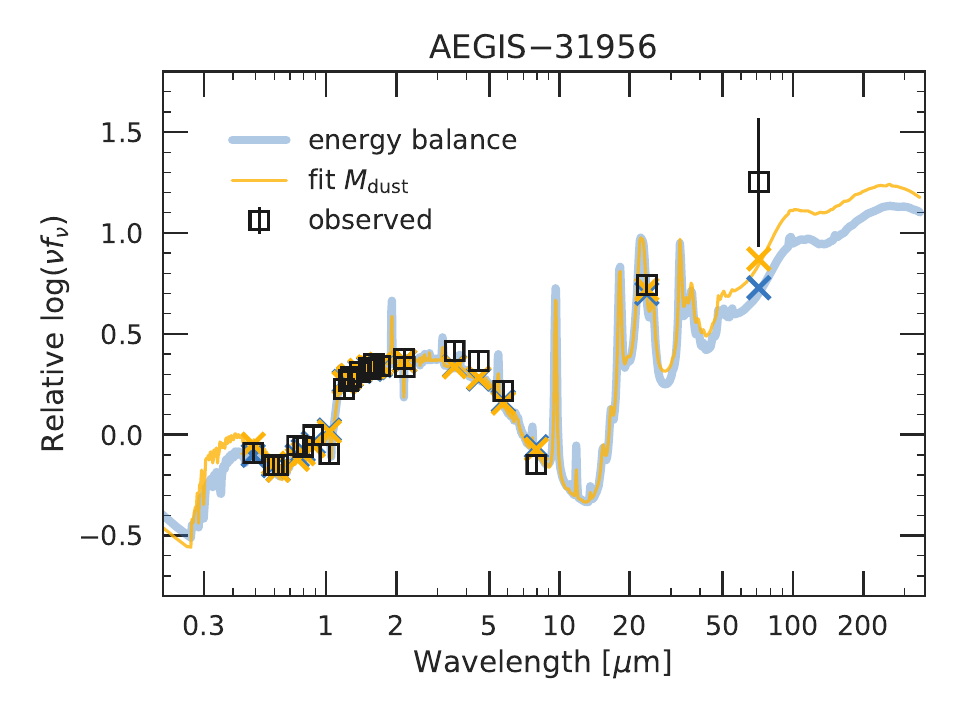}
  \caption{ An example \mcsed\ fit where mid- and far-IR data are included and the dust emission parameters are left free. The thick, blue curve reflects the SED fits when the long-wavelength normalization is estimated by assuming energy balance, while the narrow, orange curve normalizes the dust emission spectrum by leaving dust mass as a free parameter. The black squares show the observed photometric data. 
  For this object, the agreement between the two models is excellent.
  }
  \label{fig:FIRSED}
\end{figure}

The results of this experiment are shown in Figure~\ref{fig:FIRSED}.  For the galaxy AEGIS-31956, the best-fit SED using the energy balance argument is essentially identical to that generated when the long-wavelength part of the SED is fit independently of the far-UV.
As the SEDs illustrate, for this $z \sim 2$ galaxy, there are few photometric measurements in the mid- and far-IR, and those data that do exist have large uncertainties.  To take advantage of \mcsed's long-wavelength capability, one must target lower-redshift objects with better MIPS and PACs photometry which reach the peak of the IR emission.

\section{Discussion}
\label{sec:Discussion}

We have presented \mcsed, a new spectral energy distribution fitting code designed to model rest-frame UV-through-IR galaxy spectra. \mcsed\ is built for both efficiency and flexibility; the former is made possible by MCMC algorithms in the \texttt{python} package \texttt{emcee} \citep{emcee}, while the latter is achieved using several easily-adjustable prescriptions for dust attenuation, star formation rate history, stellar metallicity, and dust emission.  
The code accepts a range of observational constraints, including photometry, emission-line fluxes, and absorption-line spectral indices and is suitable for a wide range of galaxy types and redshifts.

We tested our code on a sample of $z\sim2$ emission-line galaxies identified by \citet{bowman+19} from HST grism data of the CANDELS fields.  These galaxies, which span over three orders of magnitude in stellar mass  ($M \gtrsim 10^8 \, M_{\odot}$), can be characterized as having low dust content and active star formation, and enable crucial pathfinding studies for upcoming large-scale surveys. In particular, missions such as Euclid and WFIRST will soon make such systems the dominant population of known $z>1$ galaxies. 

\begin{figure}[h!]
  \centering
  \noindent\includegraphics[width=\linewidth]{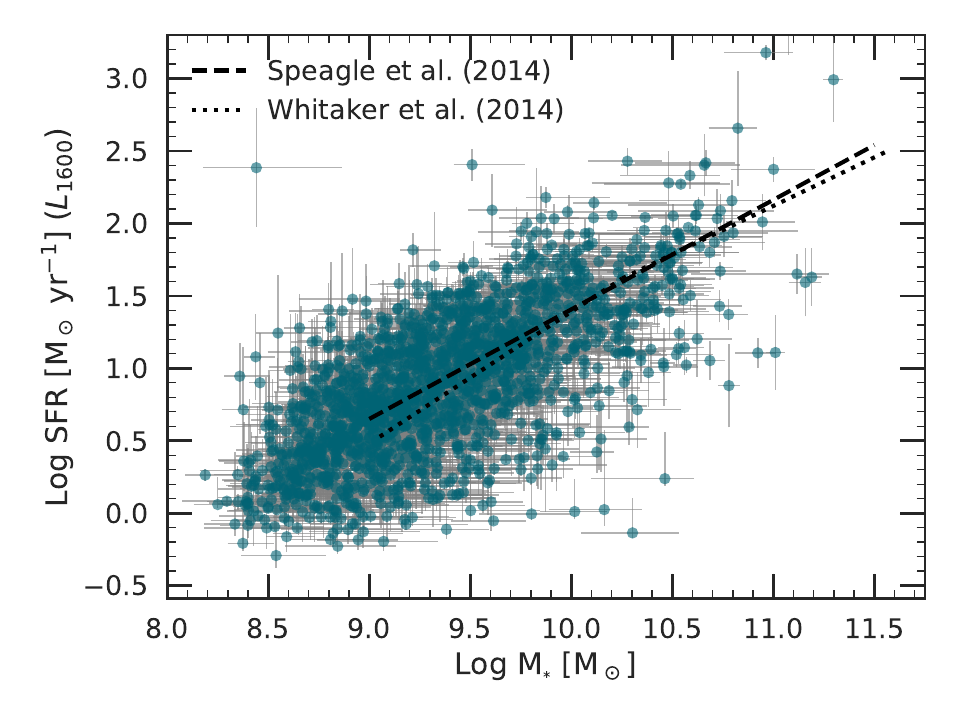}
  \caption{ The stellar mass-SFR relation for our $z\sim2$ optical emission-line galaxies. The stellar masses are estimated using the flexible, eight-parameter model described in this study, and the UV-based star-formation rates are taken from \citet{bowman+19}. The flattening in this relation that was previously seen by \citet{bowman+19} (using stellar masses estimated with a constant SFH, a \citet{calzetti+00} attenuation law, and a fixed stellar metallicity) is no longer present when more flexible fitting assumptions are adopted. For reference, two stellar mass-SFR relations from the literate are also displayed by the dashed \citep{whitaker+14} and dotted \citep{speagle+14} lines.
  }
  \label{fig:sfms_uv}
\end{figure}

The CANDELS fields, with their extensive set of observational data, are the perfect locations for investigating the physical properties of $z \sim 2$ emission-line galaxies and exploring systematic errors induced by fitting assumptions.  In this analysis, we adopted a flexible framework for our investigation, including a four age-bin star-formation rate history and a three-parameter model for dust attenuation \citep{noll+09}. We confirm that our estimates for stellar mass, recent star formation rate, and dust content are all generally consistent with the simplified fitting assumptions that appear widely in the literature. However, our investigation also suggests that biases can be introduced when using simplified assumptions: in particular, the use of a constant SFR history tends to underestimate stellar masses for systems with $M \lesssim 10^9 M_{\odot}$. Figure~\ref{fig:sfms_uv} shows the stellar mass-SFR relation using UV-based SFRs (estimated in \citealt{bowman+19}) and the updated stellar mass estimates provided by the more flexible fitting assumptions.  
Furthermore, there is compelling evidence to suggest that variations exist in the dust attenuation curve, even within our relatively homogeneous set of $z \sim 2$ emission line galaxies.  Clearly, caution should be exercised when applying a single set of assumptions to sets of galaxies with a wide range of types and redshifts.

Because most of our $z \sim 2$ emission line galaxies are quite faint, measurements of the physical properties of individual systems typically carry large uncertainties.  To circumvent this issue, we exploited the size of our sample ($\sim 2000$ galaxies) to estimate the galaxies' mean properties as a function of stellar mass.  We divided the sample into four stellar mass-bins spanning $\sim 3$~dex (from 
$10^8 \lesssim M/M_{\odot} \lesssim 10^{11}$) and computed the galaxies' mean UV-through-IR spectrum, their star formation rate histories, and dust attenuation curves.  Our emission-line galaxies exhibit an systematic shift in their SED, with the energy moving towards longer wavelengths with increasing stellar mass.  This behavior is due to both an overall increase in the dust content of the galaxies and the greater importance of older stars.  There is evidence that the shape of the attenuation curve (namely, the intrinsic greyness of the curve and the strength of the 2175~\AA\ bump) varies across the sample, though specific trends with stellar mass are marginal at best. Nonetheless, it is clear that a universal attenuation law should not be assumed across the sample, and a flexible attenuation law is more appropriate.

\acknowledgments
This study was supported through NASA Astrophysics Data Analysis grant NNX16AF33G\null.  The Institute for Gravitation and the Cosmos is supported by the Eberly College of Science and the Office of the Senior Vice President for Research at the Pennsylvania State University. This research has made use of NASA's Astrophysics Data System and the \texttt{python} packages \texttt{IPython}, \texttt{AstroPy}, \texttt{NumPy}, \texttt{SciPy}, and \texttt{matplotlib} \citep{ipython, astropy1, astropy2, numpy, scipy, matplotlib}.  This work is based on observations taken by the CANDELS Multi-Cycle Treasury Program with the NASA/ESA HST, which is operated by the Association of Universities for Research in Astronomy, Inc., under NASA contract NAS5-26555.

\facility{HST (WFC3)}

\software{\mcsed\ \citep{MCSED-code},
numpy \citep{numpy},  
astropy \citep{astropy1, astropy2}, 
emcee \citep{emcee},
scipy \citep{scipy},
matplotlib \citep{matplotlib},
corner \citep{corner},
seaborn \citep{seaborn},
dustmaps \citep{dustmaps},
logging \citep{logging}
}

\clearpage
\appendix
\section{Appendix A: \mcsed\ Description}
\label{appendix:a}

\mcsed, which is written in Python 2.7, was created with simplicity in mind.  The code is modular in nature and allows users to edit and expand the initial base to include new star formation histories, dust attenuation laws, and stellar population models. \mcsed\ is publicly available at \dataset[10.5281/zenodo.3903126]{https://doi.org/10.5281/zenodo.3903126} \citep{MCSED-code}.

The star formation histories for \mcsed\ are found in the \texttt{sfr.py} module \citep{MCSED-code}.  Included in the package are algorithms for constant and exponentially declining (or increasing) SFR histories, and a double powerlaw parameterization first proposed by \citet{behroozi+13}
\begin{equation}
{\rm SFR}(t) = A \left[\left(\frac{t}{\tau}\right)^B + \left(\frac{t}{\tau}\right)^{-C}\right]^{-1}
\label{eq:behroozi}
\end{equation}
This three-parameter formula has the flexibility to model both recent star forming activity and star formation at older epochs, and is sufficiently general to be applicable over a wide range of redshifts and stellar masses.  We also include an option which allows the user to fit an arbitrary SFR history using a table with bins in log age.   The default age bins in $\log$ years are: [$6.0 - 8.0$], [$8.0 - 8.5$], [$8.5 - 9.0$], and [$9.0 - 9.3$], adopted from \citet{prospector}. 

A number of laws are included to reproduce the extinction and attenuation of starlight due to dust.  Foreground extinction is handled via the inclusion of the  \citet{cardelli+89} relation, with the default parameter of $R_V = 3.1$.  Options to model the attenuation which occurs internal to galaxies include the starburst galaxy dust model proposed by \citet{calzetti+00}, the \citet{noll+09} generalization of the \citet{calzetti+00} law, a \citet{conroy+10} law, which is parameterized by bump strength, and the high-$z$ attenuation law of 
\citet{reddy+15} that parameterizes attenuation as a function of specific star formation rate.  These relations are found in the \texttt{dust\_abs.py} module \citep{MCSED-code}.

Strong emission lines and nebular continuum can contribute significantly to some bandpasses, and these are included in \mcsed\ via a grid of \cloudy\ models generated by \citet{byler+17}.  This emission does not necessarily see the same attenuation as the stars.  Indeed, it is well known that the light from a galaxy's \ion{H}{2} regions will be extinguished more than the light from its stars \citep[e.g.,][]{calzetti+00, battisti+16, molina+19}.  \mcsed\ handles this by allowing the user to attenuate young objects (i.e., nebular emission and stars younger than a certain age) differently from older stars.  

Finally, \mcsed\ also contains the
\citet{draine+07} prescription for dust emission.  This law is parameterized by 
the lower cutoff of the starlight intensity distribution ($U_{\rm min}$), the fraction of dust heated by starlight with $U > U_{\rm min}$ ($\gamma$), and the PAH mass fraction, $q_{PAH}$.  The total dust mass can either be directly derived from these quantities (using to assumption of energy balance) or computed via the independent normalization of far-IR data.  The code for this is located in the \texttt{dust\_emission.py} module \citep{MCSED-code}.

To create a new star formation history, dust attenuation law, or dust emission prescription, the user simply defines a new class of the following structure.  The class must initialize the parameters, define the range allowed for the parameters, and provide the MCMC initialization, \texttt{delta}.  (This last variable is equivalent to the dispersion of an initial Gaussian distribution.)  Thus, if the name of a parameter is \texttt{dummy}, the class must initialize values for \texttt{dummy}, \texttt{dummy\_lims}, and \texttt{ dummy\_delta}.  For parameters associated with star formation, the class must also include the following required functions: \texttt{set\_agelim}, \texttt{get\_params}, \texttt{get\_param\_lims}, \texttt{get\_param\_deltas}, \texttt{get\_names}, \texttt{prior}, \texttt{set\_parameters\_from\_list}, \texttt{ plot}, and \texttt{evaluate}.  The description of each of these functions can be found in one of the existing star formation history classes.  

\section{Appendix B: Mock Galaxy Tests}
\label{appendix:b}

To determine the accuracy of \mcsed , we synthesized photometry for 120 mock galaxies in each of our five fields, using the MILES spectral library, the PADOVA isochrones \citep{bertelli+94, girardi+00, marigo+08}, a \citet{kroupa01} initial mass function (IMF), a tabular SFR history of four age-bins, and a \citet{noll+09} reddening law. For each synthetic photometric measurement, we added a random error based on the image depths given by \citet{skelton+14}, and then attempted to recover the input parameters, such as the stellar mass, the coefficients for dust attenuation, and the stellar metallicity. 

\begin{figure*}
\plotone{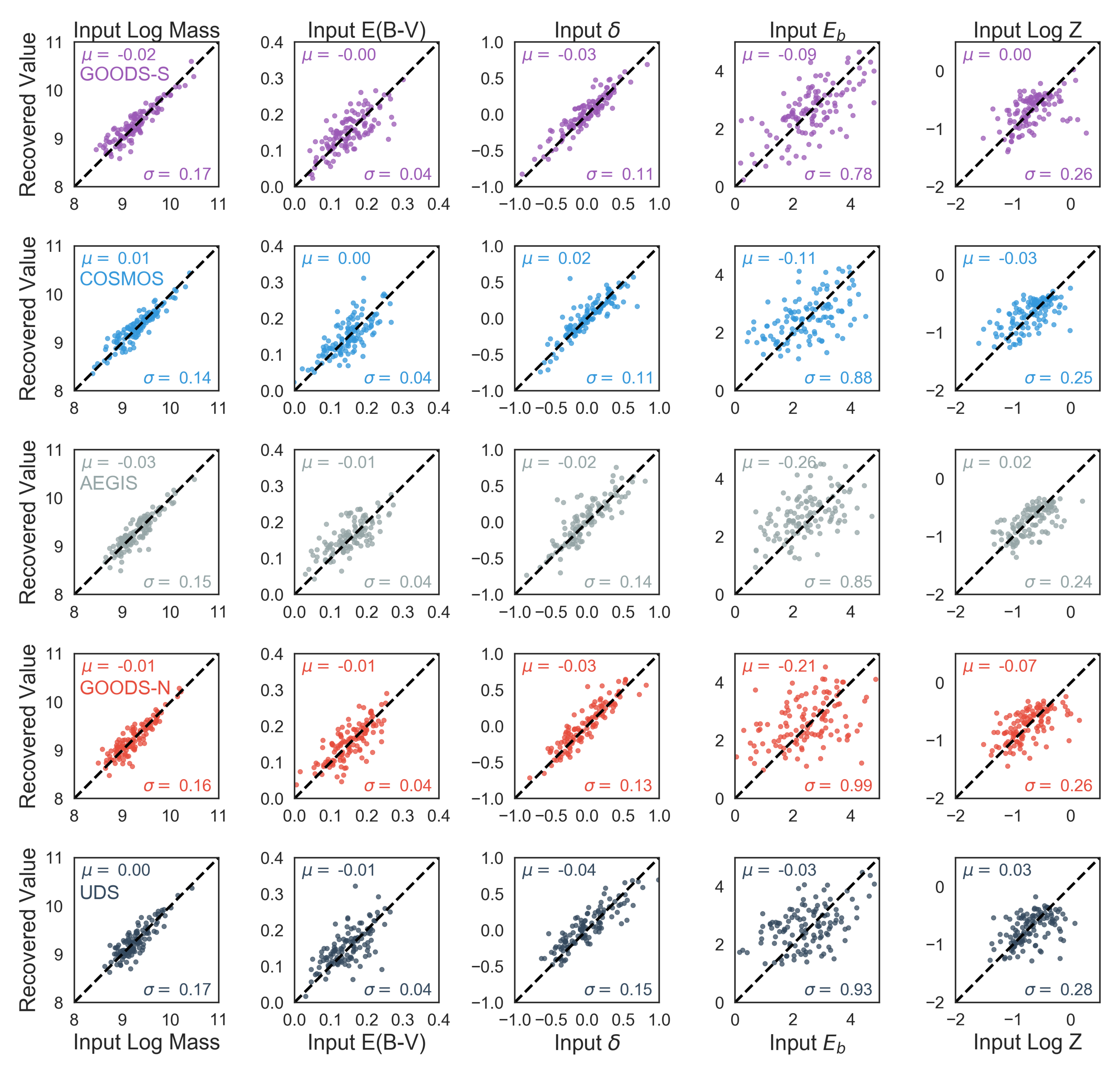}
\caption{Input SED parameters vs.~best-fit output values produced by using \mcsed\ in its test mode.  The results for each survey field are given in a row of panels.  From left to right, the columns represent the log of the stellar mass, the three dust attenuation parameters of the \citet{noll+09} dust law, and the stellar metallicity.  The mean offset and standard deviation between the input and output are noted in each panel.}
\label{fig:simulated}
\end{figure*}

The results are summarized in Figure \ref{fig:simulated}.  For each of the five fields, \mcsed\ captures the input dust attenuation law, stellar mass, and stellar metallicity quite accurately.  Specifically, the log of the stellar mass is recovered with a mean offset of $-0.01$~dex and a standard deviation of 0.16~dex, while the color excess, $E(B-V)$, is returned with a mean offset of $-0.004$ and a standard deviation of 0.04.  
\mcsed\ does a reasonably good job of inferring the power law deviation from a \citet{calzetti+00} law, $\delta$, as the mean input of the simulations is offset by $-0.03$ with a standard deviation of 0.13; however, the dust law parameters will always be difficult to estimate when the total dust content is low. More poorly modeled is the 2175~\AA\ bump, which has a small mean offset of $-0.14$ but a large standard deviation of 0.88.  This result is not surprising given the weakness of the absorption and (for the AEGIS, UDS, and GOODS-N fields) the limited amount of photometry measurements bracketing the feature.  Finally, the stellar metallicity is recovered moderately well (mean offset of $-0.01$), although with a large standard deviation of 0.26.  It is no surprise that in the GOODS-S and COSMOS fields, which have data in over 40 photometric bandpasses, produce the most accurate recovery of the input parameters.  The overall results are also quite good for the GOODS-N, AEGIS, and UDS fields.

\begin{figure*}
\plotone{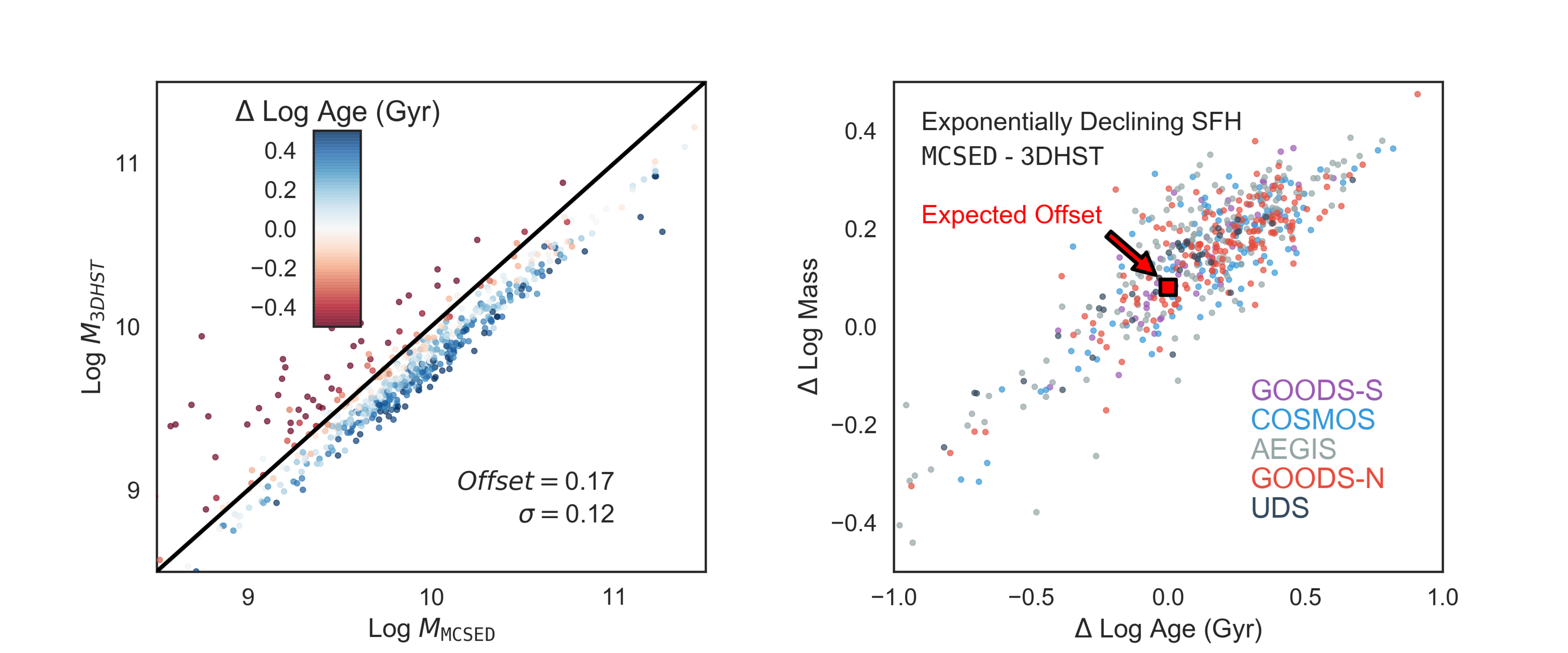}
\caption{{\it Left:} The log of the stellar mass inferred from \mcsed\ compared to that computed by \citet{skelton+14} for the galaxy sample of \citet{bowman+19}.  This comparison is for all five CANDELS fields, but includes only those sources with $z_{\rm photo}$ within 0.02 of the emission-line redshift \citep{3DHST}.  The same input parameters were used in the SED modeling, although different IMFs and SSP codes were employed.  The individual points are colored by the difference in the derived log ages (Gyr) of the systems.  {\it Right:} The difference in the log age (Gyr) versus the difference in the log stellar mass for \mcsed\ and \threedhst\null.  The expected offset in log stellar mass from differences in the IMF and SSP code is +0.08 and is shown with a red square.  There is a clear monotonic trend in the differences between the inferred ages inferred stellar masses.}
\label{fig:compare}
\end{figure*}

As an additional test, we also compared the \mcsed\ stellar masses for the \citet{bowman+19} set of $z \sim 2$ emission-line selected galaxies with those measured by the FAST code \citep{kriek+09} by \citet{skelton+14}.  For consistency, we adopted an exponentially declining star formation history, a \citet{calzetti+00} dust law, and fixed the stellar metallicity to solar.  The \citet{skelton+14} catalog used mostly photometric redshifts for calculating stellar mass, so we restricted the comparison to sources whose photometric redshifts are within 0.02 of their grism redshifts \citep[given by][]{3DHST}.  The remaining differences between the two inferred sets of models are the IMF (we used a \citealt{kroupa01} law while \citealt{skelton+14} adopted a \citealt{chabrier03} IMF), the simple stellar population code (we employed FSPS, while \citealt{skelton+14} uses \citealt{bc03}), and the general fitting methodology.  The results of the comparison are shown in Figure~\ref{fig:compare}.  The two inferred stellar mass distributions have a mean offset in log stellar mass of $+0.17$ and a standard deviation of 0.12.  In stellar mass, the systematic expected from the differing IMFs is $\sim +0.03$, while the systematic associated with the different stellar population models is $\sim +0.05$ \citep{moustakas+13, conroy13}.  The remaining difference of $+0.09$ can be accounted for by the best-fit ages which are systematically higher for \mcsed\null.  If we add a prior that the e-folding time, $\tau$, must be less than the age of the galaxy, we find excellent agreement between the two stellar masses (once the offsets from the IMF and SSP code are included).   The stellar mass comparison between \mcsed\ and \threedhst\ illustrates the systematics related to fitting methodology and the inclusion of priors.

\clearpage
\end{document}